\documentclass[twocolumn,amsmath,amssymb,prl,nofootinbib,superscriptaddress]{revtex4}

\usepackage{nicefrac, dsfont, amsmath, amsfonts, amssymb, amsthm, MnSymbol, mathrsfs, graphicx, comment, cjhebrew, times, mathtools, blkarray, xcolor, bm, notes2bib, soul}
\usepackage[all,2cell]{xy}
\usepackage[commandnameprefix=always]{changes}

\definecolor{darkblue}{rgb}{0,0,0.6}
\definecolor{darkred}{rgb}{0.6,0,0}
\definecolor{darkgreen}{rgb}{0.1,0.6,0.2}
\definecolor{darkgrey}{rgb}{0.6,0.6,0.6}
\usepackage[colorlinks=true,urlcolor=darkblue,citecolor=darkblue,linkcolor=darkred,hyperfootnotes=false]{hyperref}

\newcounter{sm}
\newcommand{\sm}[2]{
    \refstepcounter{sm}\label{#1}\vspace{.2cm} \noindent\textbf{SM\arabic{sm}. #2}\vspace{.2cm}
    }
\newcommand{\smref}[1]{SM\ref{#1}}

\newcommand{\jhat}{\bm{\hat{\jmath}}}
\newcommand{\jj}{\jmath}
\newcommand{\si}{\varsigma}
\newcommand{\ph}{\hat {\mathbf{p}}}
\newcommand{\dd}{\mathrm{d}}
\newcommand{\s}{\mathtt{s}}

\newcommand{\x}{\mathtt{x}}
\newcommand{\y}{\mathtt{y}}

\renewcommand{\r}{\boldsymbol{r}}
\newcommand{\rank}{{\operatorname{rank}}}
\newcommand{\res}{{\text{res}}}
\newcommand{\eq}{{\text{eq}}}
\newcommand{\del}{\rotatebox[origin=c]{180}{$\setminus$}}
\newcommand{\con}{\reflectbox{$\setminus$}}

\begin{document}

\title{Mutual linearity of nonequilibrium network currents}

\author{Pedro E. Harunari}
\email{pedro.harunari@uni.lu}
\affiliation{Department of Physics and Materials Science, University of Luxembourg, Campus Limpertsberg, 162a avenue de la Fa\"iencerie, L-1511 Luxembourg (G. D. Luxembourg)} 

\author{Sara Dal Cengio}
\affiliation{Laboratoire Interdisciplinaire de Physique, University of Grenoble}

\author{Vivien Lecomte}
\affiliation{Laboratoire Interdisciplinaire de Physique, University of Grenoble}

\author{Matteo Polettini}
\affiliation{Via Gaspare Nadi 4, 40139 Bologna (BO), Italy}

\begin{abstract}
For continuous-time Markov chains and open unimolecular chemical reaction networks, we prove that any two stationary currents are linearly related upon perturbations of a single edge's transition rates, arbitrarily far from equilibrium. We extend the result to non-stationary currents in the frequency domain, provide and discuss an explicit expression for the current-current susceptibility in terms of the network topology, and discuss possible generalizations. In practical scenarios, the mutual linearity relation has predictive power and can be used as a tool for inference or model proof-testing.
\end{abstract}

\maketitle

Nonequilibrium thermodynamics is usually framed as a theory of the response of observable currents to driving forces and is often predicated on its ability to describe nonlinear effects far from equilibrium, i.e.~in the absence of detailed balance. It has historical roots in such results as Einstein's relation~\cite{einstein1905motion}, Nyquist's formula~\cite{PhysRev.32.110}, the Green--Kubo and the Casimir--Onsager reciprocal relations~\cite{PhysRev.37.405, RevModPhys.17.343}, all derived under the assumption that the (mean) currents are linearly related to the driving forces. Beyond the linear regime, cornerstone results are the fluctuation relations~\cite{PhysRevLett.71.2401, PhysRevLett.78.2690}, which allow one to derive higher-order response and reciprocity relations~\cite{barbier2018microreversibility, Andrieux_2007}. Nonlinearity can lead to interesting phenomena such as relaxation slowdown~\cite{dal_cengio_geometry_2023} or negative response due to the internal activity in the system~\cite{baerts2013frenetic, falasco2019negative}, associated with complex behavior e.g.~in biological systems (homeostasis, bifurcations, limit cycles, etc.).

All these results regard the response of currents to a variation of the driving forces. However, if we take currents as the fundamental observables, it makes sense to bypass forces and establish relations among the currents themselves. This is also motivated by phenomenological considerations. Think, for example, of the mercury-in-glass thermometer once in use: it is only when the fluid stops moving that we read our body temperature, but on the other hand the thermometer scale was set by Celsius and coevals by stabilization with the universal phenomenon of heat flow between the melting ice and the boiling water at sea level~\cite{beckman1997anders, grodzinsky2020history}. Thus, the calibration of forces depends on observations about currents.

A ubiquitous framework to study fluctuations in stochastic phenomena in physics (especially at the intersection with chemistry and biology) is that of continuous-time Markov chains~\cite{gillespie_markov_1992,kampen_stochastic_2007,gardiner_stochastic_2009,norris_markov_2009}. Here, possible system configurations are represented as vertices in a network (or graph) $\mathscr{G}$  connected by edges. Transitions between vertices along an edge, in either direction, occur at rates due to the interaction of the system with the environment. Network currents then count the net number of such events, and they can be used as building blocks for all relevant thermodynamic quantities such as heat, work, entropy production, etc.: Heat flow is defined as a linear combination of network currents multiplied by the energy they displace, entropy production is a linear combination of heat flows multiplied by their conjugate thermodynamic potentials. In the long-time limit, network currents become stationary and satisfy Kirchhoff's Current Law, which is granted conservation of some underlying quantity (be it charges, matter, or, as in our case, probability). This purely topological constraint implies that not all network currents are independent. In a unicyclic network, all edges in the cycle share the same stationary current, independently of the rates. For multicyclic networks, Kirchhoff's Current Law alone does not constrain all of the currents, and since currents typically depend nonlinearly on the transition rates, there is no a priori reason to believe they should satisfy simple relations among themselves.

In fact, in this contribution we show that all stationary currents are linearly related with respect to variations of the forward and backward rates along one edge.

We consider a continuous-time Markov chain over a finite network consisting of $|\mathscr{X}|$ vertices $\x \in \mathscr{X}$ connected by $|\mathscr{E}|$ edges $e \in \mathscr{E}$, to which we assign an arbitrary orientation. We denote by $\pm e$ transitions along an edge $e$ in the direction either parallel or anti-parallel to the edge's orientation, from source vertex $\s(\pm e)$ to target vertex $\s(\mp e)$. Transitions occur at time-independent probability rates $r_{\pm e}$. The only assumption we make on the rates is that the network is irreducible, that is, that there exists a directed path of nonvanishing probability between any two vertices. In particular, the so-called cycle affinities~\cite{schnakenberg} playing the role of fundamental driving forces can take arbitrary values.

Let $p_\x(t)$ be the probability to be in state $\x$ at time $t$. Vector $\mathbf{p}(t) = (p_\x)_{\x \in \mathscr{X}}$ evolves via the master equation $\partial_t \mathbf{p}(t) = \mathbf{R} \mathbf{p}(t)$, where $\mathbf{R}$ is the rate matrix with non-diagonal elements $[\mathbf{R}]_{\s(\mp e),\s(\pm e)}= r_{\pm e}$. The normalized null vector of $\mathbf{R}$ is the unique stationary distribution $\boldsymbol{\pi}$. The stationary currents are defined as
\begin{equation}
    \jmath_e \coloneqq r_{+e} \, \pi_{\s({+e})} - r_{-e} \, \pi_{\s({-e})}. \label{eq:currents}
\end{equation}
We promote one particular edge $i$ as the input edge on the assumption it is not a bridge---an edge whose removal disconnects the graph---and study the dependence of all other stationary (output) currents on its transition rates $\r_i = (r_{+i}, r_{-i})$, while leaving all other rates unchanged. 

\begin{figure}
    \centering
    \includegraphics[width=.92\columnwidth]{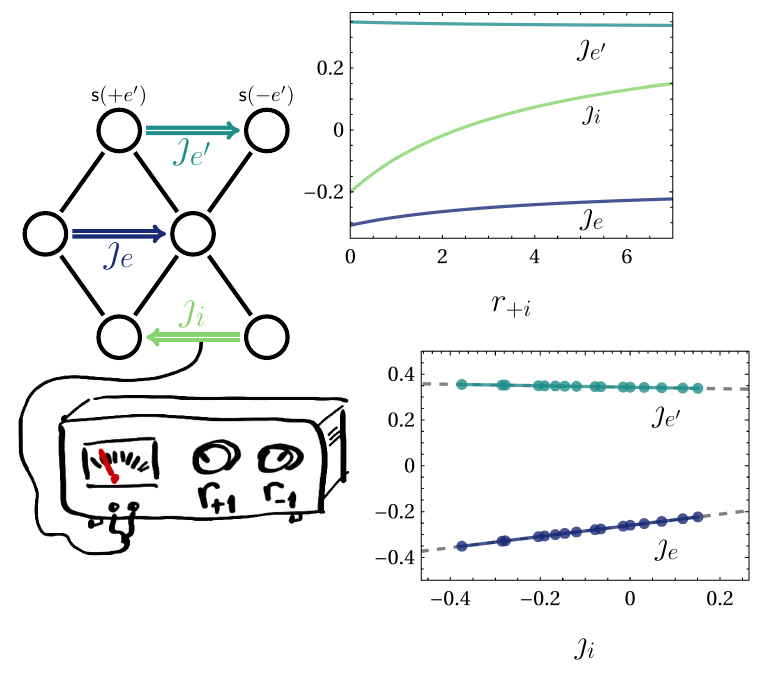}
    \caption{Scheme of the control over the input current $\jj_i$ and its linear relation to $\jj_e$ and $\jj_{e'}$ in a network. Top inset: Plot of the nonlinear relation of all three currents in terms of $r_{+i}$, with $r_{-i} = 1$. Bottom inset: plot of the two output currents' linear relation with respect to the input one, with dashed lines obtained by Eq.~\eqref{eq:result} and dots representing values of $r_{\pm i} \in [0,3]$ (see details in Sec.~\smref{app:fig-details}).}
    \label{fig:1}
\end{figure}

In general, all $\jmath_e(\r_i)$ are nonlinear functions of $\r_i$ (see Fig.~\ref{fig:1}, top inset). In fact, as a spinoff result, we prove in Sec.~\smref{app:currentbounds} that they are upper- and lower-bounded. Here, we investigate the mutual relations among the currents themselves. Inspired by Ref.~\cite{aslyamov2023nonequilibrium}, we exploit a property of the rate matrix to obtain the response of stationary currents to changes of $r_{+i}$, $r_{-i}$, or both simultaneously.
Let $\mathbf{R}_{\s(+i)}$ be defined as $\mathbf{R}$ with row $\s(+i)$ replaced by an array of ones; then the product $\mathbf{R}_{\s(+i)} \boldsymbol{\pi}$ yields a vector of zeros but for value 1 at position $\s(+i)$. This owns to the normalization $\boldsymbol{\pi}\cdot \boldsymbol{1}=1$, with $\boldsymbol{1}$ a vector with all unit entries and $\cdot$ the Euclidean scalar product. Since, in contrast to the rate matrix, $\mathbf{R}_{\s(+i)}$ is invertible (see Sec.~\smref{app:invertible} for an alternative proof to Refs.~\cite{aslyamov2023nonequilibrium, EVANS2002110}), the response of the stationary distribution can be obtained by $\partial_{r_{\pm i}} \boldsymbol{\pi} = - \mathbf{R}_{\s(+i)}^{-1} (\partial_{r_{\pm i}} \mathbf{R}_{\s(+i)}) \boldsymbol{\pi}$.
This relation can be used to obtain the responses $\partial_{r_{\pm i}} \jj_i$ and $\partial_{r_{\pm i}} \jj_e$. Their full-extent expressions can be found in Sec.~\smref{app:linearalgebra}, but the relevant piece of information is that their ratio satisfies $(\partial_{r_{\pm i}} \jj_e )/ (\partial_{r_{\pm i}} \jj_i) = \lambda^1_{e\leftarrow i}$ with $\lambda^1_{e\leftarrow i}$ independent of $\r_{i}$. Since a gradient fixes the field up to a potential, this yields the linear relation
\begin{equation}\label{eq:result}
    \jj_e(\r_{i}) = \lambda^0_{e\leftarrow i} + \lambda^1_{e\leftarrow i} \, \jj_i(\r_{i}),
\end{equation}
with $\lambda^0_{e\leftarrow i}$ also independent of $\r_{i}$. If $i$ is a bridge ($\jj_i (\mathbf{r}_i) = 0\;\forall\mathbf{r}_i $) the above formula does not hold, and $\lambda^1_{e\leftarrow i}$ diverges.

Equation~\eqref{eq:result} is our main result: control of the rates of an input edge causes a linear response in any stationary current with respect to the input one. The result is illustrated in Fig.~\ref{fig:1}. The affine coefficient $\lambda^0_{e\leftarrow i} = \jj_e (\mathbf{0})$ can easily be interpreted as the current through edge $e$ when the input rates are set to values such that the input current vanishes, a condition called stalling already shown to be relevant in traditional linear-regime theory \cite{altaner2016fluctuation}. The linear coefficient $\lambda^1_{e\leftarrow i}$ can be interpreted as a current-current edge susceptibility (from now on, simply susceptibility); we will derive and discuss an explicit expression later on.

As a generalization, consider macroscopic currents supported by many edges, $\mathcal{J}_E \coloneqq \sum_{e \in E} c_e\jj_e$ for constant coefficients $c_e$. Let $\Lambda_{E\leftarrow i}^0 = \sum_{e \in E} c_e \lambda_{e\leftarrow i}^0$ and $\Lambda^1_{E\leftarrow i} = \sum_{e \in E} c_e \lambda_{e\leftarrow i}^1$. Because
$\mathcal{J}_E(\r_{i}) = \Lambda_{E\leftarrow i}^0 + \Lambda_{E\leftarrow i}^1 \jj_i(\r_{i})$, we find that any two macroscopic currents are mutually related by
\begin{equation}\label{eq:result2_macro}
    \mathcal{J}_{E'}(\r_{i}) = \left( \Lambda^0_{E'\leftarrow i} - \frac{\Lambda_{E'\leftarrow i}^1}{ \Lambda_{E\leftarrow i}^1} \Lambda_{E\leftarrow i}^0 \right) + \frac{\Lambda^1_{{E'\leftarrow i}}}{\Lambda^1_{E\leftarrow i}} \mathcal{J}_E(\r_{i})
\end{equation}
provided $\Lambda^1_{E\leftarrow i}$ does not vanish, which can occur when all edges in $E$ are bridges. Notice that it encompasses the case of any two edge currents $\jj_{e'}$ and $\jj_e$ when $E$ and $E'$ have a single element each. For a simple illustration of the results, see the Appendix.

Mutual linearity does not extend straightforwardly to non-stationary currents, as can be checked by simple examples: In general, there do not exist time-dependent parameters $\lambda^0_{e\leftarrow i}(t)$ and $\lambda^1_{e\leftarrow i}(t)$ independent of $\r_i$ that would allow one to express $\jj_e(\r_i,t)$ as $\lambda^0_{e\leftarrow i}(t)+\lambda^1_{e\leftarrow i}(t)\jj_i(\r_i,t)$. To generalize to non-stationary currents we turn to the frequency domain. The probability distribution at time $t$ is the solution $\mathbf{p}(t) = \exp (t \mathbf{R})  \mathbf{p}(0)$ to the master equation, given an initial distribution. Defining its Laplace transform $\ph(\si)=\int_0^\infty \dd t\: e^{-\si t} \mathbf{p}(t)$ (and similarly for other functions of time), we arrive at the expression $\ph(\si) = (\si \mathbf{1}-\mathbf{R})^{-1} \mathbf{p}(0)$. Notice that both $\ph(\si)$ and the resolvent $(\si \mathbf{1}-\mathbf{R})^{-1}$ are defined for all complex numbers not in the spectrum of $\mathbf{R}$. In that domain, this allows us to obtain closed-form expressions for the derivatives $\partial_{r_{\pm i}} \hat\jj_e (\si)$ and $\partial_{r_{\pm i}} \hat\jj_i (\si)$, given in Sec.~\smref{app:laplacespace}. As in the stationary case, the important property is that their ratio $ ({\partial_{r_{\pm i}} \hat\jj_e (\si)})/({\partial_{r_{\pm i}} \hat\jj_i (\si) })$ is a constant $\hat \lambda^1_{e\leftarrow i}(\si)$ independent of $\r_i$, expressed as a ratio of cofactors of matrices related to the resolvent. We thus obtain
\begin{equation}\label{eq:result-laplace}
    \hat\jj_e(\r_{i},\si) 
    =
    \hat \lambda^0_{e\leftarrow i}(\si) 
    +
    \hat \lambda^1_{e\leftarrow i}(\si) 
    \, 
    \hat \jj_i(\r_{i},\si)
    \:.
\end{equation}
This relation generalizes the stationary result Eq.~\eqref{eq:result}, which is recovered in the small $\si$ asymptotics through
$ \lambda^0_{e\leftarrow i} = \lim_{\si\to 0} \,\si\, \hat\lambda^0_{e\leftarrow i} (\si)$
and
$     \lambda^1_{e\leftarrow i} = \lim_{\si\to 0}  \hat \lambda^1_{e\leftarrow i}(\si) $.

\begin{figure}
    \centering
    \includegraphics[width=.9\columnwidth]{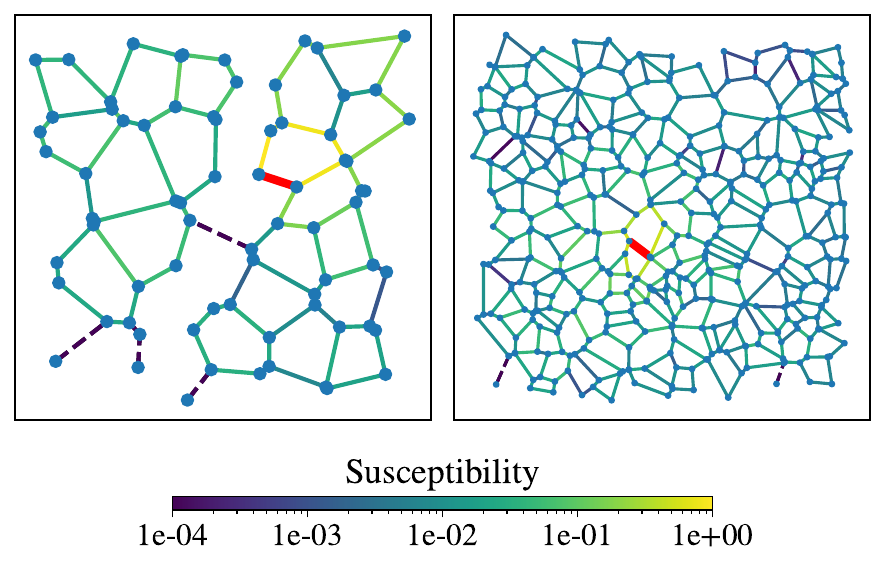}
    \caption{Illustration of the long-range interactions between currents in multicyclic networks. We built the networks from Voronoi diagram repartitions of the plane and we randomized the rates between $1/2$ and $1$. The susceptibilities are computed using Eq.~(\ref{eq:lambda1_explicit}) for every edge with respect to the input edge (thick red edge). Here we plot the absolute value of the susceptibilities. Zero susceptibilities are represented with dashed lines and correspond to bridges. Left: the perturbation crosses the bridge and affects all the network currents. This long-range effect of single-edge perturbation is absent in detailed-balance networks. Right: the susceptibilities decrease at large distance but never vanish (except for bridges leading to leaves). The susceptibilities present a degree of heterogeneity at large distance with patches or single edges where the response is screened. See Sec.~\smref{app:fig-details} for full details.}
    \label{fig:3}
\end{figure}

Furthermore, the Laplace formalism allows us to obtain an explicit expression for $\lambda^1_{e\leftarrow i}$ in terms of sums over rooted spanning trees. We recall that in an oriented graph, a spanning tree $\mathscr{T}_{\x}$ with root $\x$ is a subset of edges such that every vertex of the network is connected to $\x$ via a unique path and every edge along such path points towards $\x$. It is well-known that, up to normalization, the stationary distribution can be written as $\pi_\x \propto \tau_\x$~\cite{hill_studies_1966, schnakenberg, avanzini_methods_2023}, where  $\tau_\x = \sum_{\mathscr{T}_{\x}\subseteq\mathscr G} w(\mathscr{T}_{\x})$
is the spanning-tree polynomial, namely the sum over rooted spanning trees $\mathscr{T}_{\x}$ of the product $w(\mathscr{T}_{\x})$ of the transition rates along the tree. This result is termed the Markov chain tree theorem and is valid for arbitrary transition rates (it has been used recently~\cite{liang2023universal,arunachalam2024information} to derive bounds in such models).
It is thus natural to use spanning-tree ensembles to represent the susceptibility $\lambda^1_{e\leftarrow i}$, but in this case we need to expand on these concepts. In particular, we borrow the notation $\del$ and $\con$ from the deletion-contraction paradigm of undirected graphs. We define $
\tau_{\del i} = \sum_\x  \sum_{\mathscr T_\x\subseteq \mathscr G, i \not\subseteq  \mathscr T_\x} w\big(\mathscr{T}_{\x}\big)$ as the sum over the subset of trees $\mathscr{T}_{\x}$ spanning $\mathscr G$ which do not contain edge $i$. We also define $\tau^{\con i} = \sum_\x  \sum_{\mathscr T_\x\subseteq \mathscr G, i \subseteq  \mathscr T_\x} w\big(\mathscr{T}_{\x} \del i \big)$ as the sum over the subset of trees $\mathscr{T}_{\x}$ containing edge $i$ of the product $w\big(\mathscr{T}_{\x} \del i \big)$ of all rates but that of $i$. We say that edge $i$ is, respectively, deleted or contracted in the two operations. Notice that $\tau_{\del i}$ and $\tau^{\con i}$ are both independent of $\r _i$. 
Finally, we obtain for the susceptibility (see Sec.~\smref{app:graphrepresentation}):
\begin{align}
    \lambda^1_{e\leftarrow i}
    \ =\ & \,
    r_{+e}\,\frac
    {
    \displaystyle
    \tau_{\del i, e}^{\con \s(+e)\to \s(+i)}-\tau_{\del i, e}^{\con \s(+e)\to \s(-i)}
    } 
    {\tau_{\setminus i}}
\nonumber
\\[1mm]
 &  \
    - 
    r_{-e}\,
    \frac{
    \displaystyle
    \tau_{\del i, e}^{\con \s(-e)\to \s(+i)}-\tau_{\del i, e}^{\con \s(-e)\to \s(-i)}
    }
    {\tau_{\setminus i}}
 \:.
\label{eq:lambda1_explicit}
\end{align}
Each term in the numerator of Eq.~(\ref{eq:lambda1_explicit}) corresponds to the spanning tree polynomial of a modified network built from $\mathscr{G}$ by, first, removing edges $i$ and $e$ and, second, adding and contracting a directed edge from $\s(\pm e)$ to $\s(\pm i)$ (see Fig.~\ref{fig:2} for an illustration). This means that the correct spanning-tree ensemble to compute the susceptibility is that of the original network $\mathscr G$ deprived of both edges $e$ and $i$ where one connects the vertices of the input and output edges by adding a directed edge from $\s(\pm e)$  to $\s(\pm i)$. This operation of connection is non-local, giving rise to long-distance interactions between currents (see Fig.~\ref{fig:3}).

The susceptibility depends on kinetic and topological properties of the process (as is the case for the bounds for state observables proven in Refs.\:\cite{aslyamov2023nonequilibrium, owen2020universal, PhysRevE.108.044113, aslyamovGeneralTheoryStatic2024}). The form of Eq.~\eqref{eq:lambda1_explicit} implies that the susceptibility is a monotonic function of every $r_{\pm e'}$ (with $e'\neq i$) and is invariant by a global rescaling of the rates. Its extrema are thus reached by setting rates to $0$ or $+1$, corresponding to ``skeleton'' networks that maximize or minimize the influence of the input current to the output one. In particular, notice that if $i$ is a bridge, $\tau_{\setminus i}$ vanishes (as there are no spanning trees not containing edge $i$) and the susceptibility is ill-defined; in fact in that case, the input current is zero independently of $\r_i$. Interestingly, though, Eq.~\eqref{eq:lambda1_explicit} implies that in networks that have a bridge, the susceptibility does not vanish even when the input and output currents are on opposite sides of the bridge, despite the susceptibility (and the current) of the bridge being zero. This is due to the dependency of $\boldsymbol{\pi}$ in all the rates, out of equilibrium (see Fig.~\ref{fig:3}). Thus, Eq.~\eqref{eq:lambda1_explicit} expresses how controlling the current of edge $i$ builds long-distance interactions with other currents, which may be related to the overall activity~\cite{maes2020frenesy} of the system. An additional result regarding bridges is that all currents are strictly linear one to another (without affine coefficient) when they live on a different island than the input edge (see Sec.~\smref{app:bridge}).

The Markov-chain network formalism is intimately connected to the description of deterministic unimolecular chemical reaction networks (CRNs) with mass action law (see e.g.~\cite{schnakenberg, HillBook, hill_studies_1966, clarke_stoichiometric_1988, feinberg2019foundations, PhysRevX.6.041064}).
A vertex $\x\in\mathscr X$ represents a chemical species $A_\x$ and 
an edge $e\in\mathscr E$ a bidirectional reaction
$A_{\s(+e)} \rightleftarrows A_{\s(-e)} $ 
occurring at rate $r_{+e}$ (resp.~$r_{-e}$) in the forward (resp.~backward) direction.
The vector $\mathbf{p}(t)$ of species concentrations 
also evolves through  $\partial_t \mathbf{p}(t) = \mathbf{R}\, \mathbf{p}(t)$.
In such settings, which describe closed (i.e.~non-chemostatted) CRNs,
the results we have described so far are translated in a direct manner:
the system reaches stationarity at large times,
and, upon controlling $\jj_i$ through $\boldsymbol{r}_i$, output currents $\jj_e$ satisfy the linearity relation Eq.~\eqref{eq:result}.
The sole difference is that,
$\sum_\x p_\x(t)$ being conserved,
the normalization of the stationary concentration $\boldsymbol{\pi}$
is fixed by its initial value $\mathbf{p}(0)$
through $\boldsymbol{\pi}\cdot \boldsymbol{1} = \mathbf{p(0)}\cdot \boldsymbol{1}$
(assumed to be independent of the rates).

We now show that the mutual linearity of currents can be extended 
to the case of open (i.e.~chemostatted) unimolecular CRNs.
To do so, we drive the system by chemostatting 
a subset of species $\y\in\mathscr Y \subseteq \mathscr X$:
reservoirs
create or destroy these species through reactions $\emptyset \rightleftarrows A_\y$,
with given rates.
As shown in~\cite{dal_cengio_geometry_2023}, 
it is useful to represent such a drive by adding $|\mathscr Y|$ edges $f\in\mathscr F$,
each directed from a single new vertex $\emptyset$
to a chemostated species $\y = \s(-f)$.
The stationary currents of the corresponding reactions are
\begin{equation}
    \label{eq:jres}
    \jj_f = r_{+f} - r_{-f} \pi_{\s(-f)}
\end{equation}
where $r_{+f}$ (resp.~$r_{-f}$) is the creation (resp.~destruction) rate
of species $\y=\s(-f)$.
Importantly, 
such currents are affine functions of the stationary concentration $\boldsymbol{\pi}$,
in contrast to Eq.~\eqref{eq:currents}.
The same holds for the time-dependent current, implying
that the total concentration is not preserved (the dynamics is not conservative).
However, one can obtain $\boldsymbol{\pi}$ by mapping the open system to a closed linear system,
as follows.
We consider a closed CRN on a
graph of vertices $\{\emptyset\}\cup\mathscr X$
and edges $\mathscr E \cup\, \mathscr F$,
and denote by $\mathbf{R}^\res$ its rate matrix.
Its stationary concentration is 
a $(|\mathscr X|+1)$-dimensional vector $\boldsymbol{\pi}^\res$ 
solution of $\mathbf{R}^\res\,\boldsymbol{\pi}^\res=\boldsymbol{0}$, 
that we normalize by imposing $\pi^\res_\emptyset=1$.
This condition, see Eq.~\eqref{eq:jres}, ensures that its stationary currents
are identical to that of the open CRN above; by unicity, 
we thus have $\pi_\x=\pi^\res_\x$ for $\x\in\mathscr X$~\bibnote{%
Notice that this implies that the stationary concentration of such open CRNs
can be expressed using the Markov chain tree theorem on the rate matrix $\mathbf{R}^\res$.
}.
Since the normalization $\pi^\res_\emptyset=1$ 
imposes a rates-dependent constraint,
the derivation of the mutual linearity has to be modified~\bibnote{%
The normalization  $\pi^\res_\emptyset=1$ 
makes that $\boldsymbol{\pi}^\res\cdot \boldsymbol{1}$ is different from $1$
and depends generally on all the rates (as seen from the Markov chain tree theorem).
In particular we have  that $\mathbf{R}^\res_\x \boldsymbol{\pi}^\res\neq \boldsymbol{\delta}_\x$: $\mathbf{R}^\res_\x \boldsymbol{\pi}^\res$ actually depends on the rates,
so that the proof used in Markov chains would not apply.
}.
We proceed as follows:
defining $\bar{\mathbf{R}}^\res_{\x}$ 
by replacing  line $\x$ of ${\mathbf{R}}^\res$ by $\boldsymbol{\delta}_\emptyset\equiv (1,0\ldots 0)$ 
[placing species $\emptyset$ first],
the stationarity condition  
$\mathbf{R}^\res \boldsymbol{\pi}^\res=\boldsymbol{0}$
implies
$\bar{\mathbf{R}}^\res_\x \boldsymbol{\pi}^\res =\boldsymbol{\delta}_\x$ 
(Kronecker delta vector for vertex $\x$).
Using then the invertibility of $\bar{\mathbf{R}}^\res_\x$ (see Sec.~\smref{app:invertible}),
we express $\partial_{r_{\pm i}} \boldsymbol{\pi}^\res$
using $\bar{\mathbf{R}}^\res_{\s(+i)}$ and its inverse.
As in the Markov-chain case,
this yields that the ratio  
$(\partial_{r_{\pm i}} \jj_e )/ (\partial_{r_{\pm i}} \jj_i) $
is independent on the rates $\boldsymbol{r}_i$,
and allows one to conclude that the mutual linearity 
of Eq.~\eqref{eq:result} holds for open CRNs (see Sec.~\smref{app:linearalgebra} for details). Noteworthy, a chemostatting current $\jj_f$ can be the input or output current (if two or more species are chemostatted, ensuring $f$ is not a bridge).

Let us now draw conclusions and discuss open questions.

We have already seen that linearity is not a simple consequence of Kirchhoff's Current Law. Neither it is a straightforward consequence of the spanning-tree expression for the stationary distribution, by replacement of $\pi_{\s(\pm e)}$ in Eq.~\eqref{eq:currents}. We will explore in a forthcoming contribution some more spanning-tree combinatorics related to our main result.

The main strength of our result is that, from an operational perspective, two measurements of two currents suffice to determine $\Lambda_{E\leftarrow i}^0$ and $\Lambda_{E \leftarrow i}^1$, so further measurements have predictive power. Furthermore, the result holds in networks with more than one edge between a pair of states, and in networks with unidirectional transitions (absolute irreversibility) which typically pose a thermodynamic conundrum~\cite{murashita2014nonequilibrium, baiesi2023effective}.When applied to (open) resistor networks,
where $r_{+e}=r_{-e}$ is the resistance of edge $e$,
our result retrieves the ``principle of superposition'' of linear electric networks (see e.g.~Chap.~5 of~\cite{seshu_linear_1959}).

Although the main limitation of our result is the assumption that only the forward and backward rates of one specific transition are varied, this is met in several Markov-based biophysical models of molecular motors~\cite{bierbaum_chemomechanical_2011, Verbrugge2007, chemla2008exact}, conformational dynamics~\cite{chodera_markov_2014, suarez2021markov, malmstrom2014application}, DNA transcription~\cite{abbondanzieriDirectObservationBasepair2005}, kinetic proofreading~\cite{yuEnergyCostOptimal2022, banerjeeElucidatingInterplaySpeed2017}, and other processes~\cite{allen2010, floydLearningControlNonequilibrium2024}, where rates along a single edge might be controlled by changing the concentration of a reactant chemical species (e.g.~an enzyme, on the assumption of enzyme specificity). More concretely, consider an established model for the molecular motor Myosin-V~\cite{skauKineticModelDescribing2006}; perturbations in the concentration of inorganic phosphate yield a linear relation between ATP consumption and the motor velocity, with affine coefficient reflecting the consumption of ATP when the motor stalls. See the Appendix for more details.

Another area of future investigation is whether the result eventually extends to population dynamics, e.g.~stochastic chemical reaction networks and shot-noise electronic devices~\cite{freitas2021stochastic} where the network is potentially unbounded and the same parameter affects an infinite number of network transitions. As regards open networks of interacting units, the concept of susceptibility in interacting transport (e.g.~vehicular) systems has been studied in Ref.\cite{rolando2023failure}.

In some physical systems, transition rates are parametrized according to 
local detailed balance~\cite{PhysRevE.85.041125, falascoLocalDetailedBalance2021a}, e.g.~$ r_{+i}/r_{-i} = \exp \{ \beta_i [ \epsilon_{\s(+i)} - \epsilon_{\s(-i)} ]\}$ with $\beta_i$ the inverse temperature of a reservoir and $\epsilon_\x$ the energy of state $\x$. Our results apply for instance when varying $\beta_i$ on a single edge. In fact, it was found that perturbing the energy of a single vertex (thus modifying the rates of all of its outward transitions) leads to a constant ratio between any currents~\cite{mallory_kinetic_2020}. Another example of edge perturbation is the change of a kinetic barrier between two states.

An interesting area of overlap and future inspection is the interplay of our result with recently proposed frameworks for the composition of nonlinear chemical reaction networks~\cite{avanzini2023circuit} or of generic thermodynamic devices~\cite{raux2023circuits}, extending concepts from linear electrical circuit theory such as that of the conductance matrix. Interestingly, however, we could not find any immediate connection of our result to the usual machinery of response theory or of large deviations, fluctuation relations, and the like. This could be an interesting area of inspection, in particular as it comes to figures of merit such as efficiency and the quality factor, which relate input and output currents to benchmark performance and allow exploration of regimes and limits of operation.

Another possibility is to use our results to make inferences about the topology and rates of the underlying network. For example, detecting nonequilibrium from available observables is relevant in many fields, in particular biophysics~\cite{fangNonequilibriumPhysicsBiology2019, gnesottoBrokenDetailedBalance2018, zollerEukaryoticGeneRegulation2022, hartichNonequilibriumSensingIts2015, yangPhysicalBioenergeticsEnergy2021, harunari2024unveiling}. As proven in Sec.~\smref{app:mutual-positivity}, if the signs of susceptibilities are non-reciprocal upon swapping input and output edges, $\lambda_{e \leftarrow i}^1 / \lambda_{i \leftarrow e}^1 < 0$, the network is out of equilibrium (non-reciprocal edge perturbations thus require dissipation). Similarly, networks satisfying detailed balance will have zero susceptibility in all edges separated from the input by a bridge (see Sec.~\smref{app:bridge}). The coefficients $\lambda^0$ and $\lambda^1$ can be empirically obtained and compared to theoretical predictions of a candidate model using Eq.~\eqref{eq:lambda1_explicit} [or alternatively Eqs.~\eqref{eq:ratio} and \eqref{eq:lambda1limratiodet}]. Further inference schemes might arise from inspecting how susceptibilities change along cycles or decay with a notion of distance.

\appendix*
\section{Appendix: Encompassing example of a simple molecular motor}\label{appendix}

\begin{figure}
    \centering
    \includegraphics[width=.85\columnwidth]{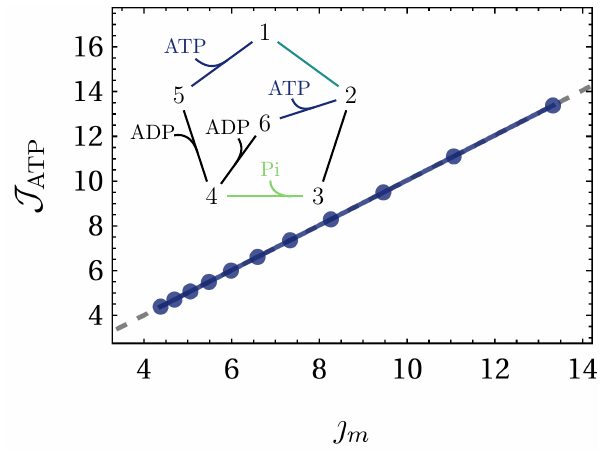}
    \caption{Dots represent the current of ATP consumption and mechanical movement for perturbed values of [Pi], the dashed gray line is obtained from Eq.~\eqref{eq:Jatpjm}. Inset: Network of the Myosin-V model.  Further details in Sec.~\smref{app:fig-details}.}
    \label{fig:myosin_model}
\end{figure}

To illustrate our results, we consider a Markov model that describes how the Myosin-V protein moves along an actin filament fueled by the consumption of ATP \cite{skauKineticModelDescribing2006, mallory_kinetic_2020}, see Fig.~\ref{fig:myosin_model}. Each state represents a configuration of the two protein heads according to their attachment to the filament. Some transitions consume/release ATP, ADP or Pi, which is often undetected by experiments, while the flux along states 1 and 2 represents the mechanical movement. In this model, the main cycle (12345) represents the net movement of the motor, while cycle (2346) represents the futile consumption of ATP that occurs when the front head detaches and reattaches to the filament without displacement.

An analysis using Kirchhoff's current law reveals that the stationary consumption rate of ATP $\mathcal{J}_\text{ATP} \equiv \jj_{5\to 1} + \jj_{6\to 2}$ equals both the stationary release rate of ADP, $\jj_{4\to 5} + \jj_{4\to 6}$, and of Pi, $\jj_\text{Pi} \equiv \jj_{3\to 4}$. The net movement of the protein is given by $\jj_\text{m} \equiv \jj_{1 \to 2}$ and, again due to Kirchhoff, satisfies
\begin{equation}\label{eq:kirchhoff}
    \mathcal{J}_\text{ATP} = \jj_{6\to 2} + \jj_\text{m}\,.
\end{equation}
These relations hold regardless of any perturbations since they are topological constraints. We now consider changes in the environment availability of inorganic phosphate, Pi, whose concentration enters as $r_{4\to 3} \propto \text{[Pi]}$. Varying [Pi] thus represents a single rate perturbation that yields a nonlinear change in all fluxes. Using our main result Eq.~\eqref{eq:result}, $\jj_\text{m}$ is found to be linearly related to the flux of Pi, $\jj_\text{m} = \lambda^0_{\text{m} \leftarrow \text{Pi}} + \lambda^1_{\text{m} \leftarrow \text{Pi}} \jj_\text{Pi}$,
where we adopt the notation that $\lambda^{0,1}_{\text{m} \leftarrow \text{Pi}}$ are the coefficients related to the perturbation of [Pi] that can be obtained empirically or using their analytical expressions. A similar result holds for the ATP consumption in the main cycle $\jj_{5\to 1}$ and in the futile cycle $\jj_{6\to 2}$. 

Now, using the generalization for ``macroscopic'' currents Eq.~\eqref{eq:result2_macro}, we find that $\mathcal{J}_{\rm ATP}$ and $\jj_\text{m}$ satisfy themselves a linear relation regardless of the futile consumption $\jj_{6\to 2}$ upon perturbations of [Pi]:
\begin{align}\label{eq:Jatpjm}
    \mathcal{J}_\text{ATP} =& \lambda^0_{(5 \to 1) \leftarrow \text{Pi}} + \lambda^0_{(6 \to 2) \leftarrow \text{Pi}} - \frac{\lambda^1_{(5 \to 1) \leftarrow \text{Pi}} + \lambda^1_{(6 \to 2) \leftarrow \text{Pi}}}{\lambda^1_{\text{m}\leftarrow \text{Pi}}} \lambda^0_{\text{m} \leftarrow \text{Pi}} \nonumber\\
    &+ \frac{\lambda^1_{(5 \to 1) \leftarrow \text{Pi}} + \lambda^1_{(6 \to 2) \leftarrow \text{Pi}}}{\lambda^1_{\text{m}\leftarrow \text{Pi}}} \jj_\text{m}
\end{align}
where, importantly, none of the coefficients depend on the concentration of Pi, see Fig.~\ref{fig:myosin_model}. Hence, results of the present manuscript allow one to derive a relation between ATP and mechanical currents which, instead of depending on an unknown current $\jj_{6\to 2}$ (see Eq.~\eqref{eq:kirchhoff}), is affine with coefficients that, interestingly, are independent of [Pi]. It establishes a direct relationship, robust to perturbations of [Pi]. The affine coefficient in Eq.~\eqref{eq:Jatpjm} represents the consumption of ATP when the motor stalls, which is non-zero in our settings. Also, notice that the velocity of the motor is $\jj_\text{m}$ times the step size, and is also linear with the total ATP consumption.

\begin{figure}
    \centering
    \includegraphics[width=.8\columnwidth]{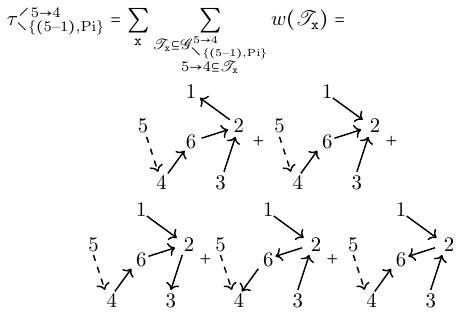}
    \caption{Evaluation of $\tau_{\del \{ (5\text{--}1), \text{Pi} \} }^{\con 5 \to 4}$. Solid arrows represent the transition rates multiplied to form the polynomial of each tree, and the dashed arrows represent the contracted edge that is not included in the polynomial.}
    \label{fig:myosin_trees}
\end{figure}

In this contribution, we provide more than one approach to obtain the susceptibilities involved in Eq.~\eqref{eq:Jatpjm}: $(i)$ empirically (collecting two data points of current vs. current), $(ii)$ from determinants [see Eqs.~\eqref{eq:ratio} and \eqref{eq:lambda1limratiodet}], or $(iii)$ using spanning trees [Eq.~\eqref{eq:lambda1_explicit}]. To illustrate the latter, consider the susceptibility $\lambda^1_{(5\to 1)\leftarrow \text{Pi}}$. From Eq.~\eqref{eq:lambda1_explicit}, we need to evaluate the polynomials of rooted spanning trees of modified networks, one of them is $\tau_{\del (5\text{--}1), \text{Pi}}^{\con 5 \to 4}$, which corresponds to the spanning trees of the original network deprived of the input edge $\text{Pi} =(3\text{--}4)$ and from the output edge $(5\text{--}1)$, and including the connecting edge $5 \to 4$. All the rooted spanning trees of this network with the contraction of $5 \to 4$ are represented in Fig.~\ref{fig:myosin_trees}.

An analysis of the coefficients' numerical values is also informative. Using the experimentally motivated transition rates described in \cite{mallory_kinetic_2020}, we find that the affine part of Eq.~\eqref{eq:Jatpjm} is $8\times 10^{-11}$, while the susceptibility is $1.005$. It tells that the rates were selected so that the futile consumption of ATP is small and most ATP consumption is directly transformed into movement.

\begin{acknowledgments}

\textit{Data Availability Statement:} A comprehensive tutorial covering the main ideas and codes to generate the figures are available in the public repository~\cite{github}.

\textit{Acknowledgments:} We thank Timur Aslyamov and Qiwei Yu for fruitful discussions. The research was supported by the National Research Fund Luxembourg (project CORE ThermoComp C17/MS/11696700), by the European Research Council, project NanoThermo (ERC-2015-CoG Agreement No. 681456), and by the project INTER/FNRS/20/15074473 funded by F.R.S.-FNRS (Belgium) and FNR (Luxembourg). SDC and VL acknowledge support from IXXI, CNRS MITI and the ANR-18-CE30-0028-01 grant LABS.
\end{acknowledgments}

\bibliographystyle{ieeetr}
\bibliography{biblio}

\newpage

\onecolumngrid
\setcounter{subsection}{0}
\renewcommand{\thesubsection}{SM\arabic{subsection}}
\setcounter{equation}{0}
\renewcommand{\theequation}{S\arabic{equation}}

\vspace{.2cm}
\begin{center}
    \textbf{Supplemental Material}
\end{center}
\vspace{.2cm}

In Sections~\smref{app:currentbounds}-\smref{app:graphrepresentation}, we assume without loss of generality that the input edge is $i=1$.

%------------------------------------------------
\sm{app:currentbounds}{Current bounds}
%------------------------------------------------

When a current is embedded in a network and its removal preserves irreducibility, it cannot assume any possible value upon perturbation of its rates, since the paths through the rest of the network will form a bottleneck. Intuitively, if the rate in the convention-defined direction of a current is very large while the opposite is small, the system will rapidly flow through this edge, but to return to its source, the system will have to take a detour through the network, rendering the current upper bounded by the topology and all other rates.

The input current \(\jj_1\) is monotonically increasing in terms of $r_{+1}$ and decreasing in terms of $r_{-1}$. It means that its extrema are located at the limit of infinite rates, consistent with the intuition above.

In a graph, a rooted spanning tree $\mathscr{T}_\x$ with root $\x$ is a spanning tree such that each edge is directed along the unique path that leads to the root. Let $w(\cdot)$ be the function that takes the product of all rates in a subset of edges. By the Markov chain tree theorem, the stationary probability of a state $\x$ is given by
\begin{equation}
    \pi_\x = \frac{\tau_\x}{\sum_\y \tau_\y},
\end{equation}
where $\tau_\x \coloneqq \sum_{\mathscr{T}_\x}  w(\mathscr{T}_\x)$ is the rooted spanning tree polynomial with its sum spanning through all possible rooted spanning trees. We adopt the notation $\tau_\x = \tau_{\x \; \del \pm 1} + r_{+1} \tau_\x^{\con + 1} + r_{-1} \tau_\x^{\con - 1}$, where the first term $\tau_{\x \; \del \pm 1}$ accounts for the spanning trees that do not contain edges $\pm 1$, $\tau_\x^{\con + 1}$ for the trees that contain the edge $+1$ but its rate is removed from the polynomial, and analogously for $\tau_\x^{\con - 1}$. Notice that it is not possible to have both input rates in the same spanning tree by the definition of a tree.

The input current can be expressed as
\begin{equation}
    \jj_1 = \frac{r_{+1} \tau_{\s(+1) \; \del -1} - r_{-1} \tau_{\s(-1) \; \del +1} }{\sum_\x \tau_{\x \; \del \pm 1} + r_{+1} \tau_{\x \; \del - 1}^{\con +1} + r_{-1} \tau_{\x \; \del + 1}^{\con -1} },
\end{equation}
and therefore it is bounded by
\begin{equation}
    \jj_1^\text{min} \coloneqq \lim_{r_{-1} \to \infty} \jj_1  = - \frac{\tau_{\s (-1) \; \del +1} }{ \sum_\x \tau_{\x \; \del +1}^{\con -1} }
\end{equation}
and
\begin{equation}
    \jj_1^\text{max} \coloneqq \lim_{r_{+1} \to \infty} \jj_1  = \frac{\tau_{\s (+1)\; \del -1} }{ \sum_\x \tau_{\x \; \del -1}^{\con +1} }.
\end{equation}
Both bounds are finite when the transition rates of the network are also finite since $\sum_\x \tau_{\x \; \del +1}^{\con -1} > 0$ and $\sum_\x \tau_{\x \; \del -1}^{\con +1} > 0$. As a sanity check, in the case of a trivial cycle-free system, the numerators vanish and the bounds collapse to $\jj_1 =0$. For the case of a single cycle, the bounds change upon affinity-preserving transformation, indicating that the cycle affinity itself is not sufficient to predict minimal and maximal currents with respect to single-edge perturbations.

If there exists a spanning tree rooted at $\s(-1)$ with nonzero rates and not containing $+1$, i.e.~if $\tau_{\s (-1) \; \del +1} >0$, the input current can take negative values. Similarly, the input current can be positive if there is a spanning tree rooted at $\s(+1)$ with nonzero rates and not containing $-1$. Upon adding edge $\pm 1$ to these trees, a cycle is formed, which means that if the current belongs to a cycle, it can always assume positive, zero, and negative values just by tuning its transition rates.

%------------------------------------------------
\sm{app:invertible}{Invertibility of the auxiliary matrices \texorpdfstring{$\mathbf{R}_\x$}{Rx} 
and \texorpdfstring{$\bar{\mathbf{R}}^\res_\x$}
{Rresx}}
%------------------------------------------------

As in the main text, we define $\mathbf{R}_\x$ from the rate matrix $\mathbf{R}$ by replacing its line corresponding to vertex $\x$ by a line of ones.
We present an alternative proof for the invertibility of $\mathbf{R}_\x$ with respect to those of Refs.~\cite{aslyamov2023nonequilibrium, EVANS2002110}. The continuous-time Markov chain generator $\mathbf{R}$ of an ergodic process has a unique eigenvector, the stationary probability $\boldsymbol{\pi}$, and therefore $\mathrm{dim}\,\mathrm{ker}\,\mathbf{R} = 1$. Thus, by the rank-nullity theorem, $\mathrm{rank}\,\mathbf{R} = |\mathscr{X}| - 1$. Since, by the Perron--Frobenius theorem, the vector of ones $\boldsymbol{1}$ cannot be orthogonal to the kernel of $\mathbf{R}$---that is, $\boldsymbol{1}\cdot\boldsymbol{\pi}\neq 0$---it cannot be in the coimage of $\mathbf{R}$ and therefore cannot be obtained as a linear combination of the rows of $\mathbf{R}$. Therefore, the union of any $|\mathscr{X}|-1$ rows of $\mathbf{R}$ and the vector $\boldsymbol{1}$ span a $|\mathscr{X}|$-dimensional space, rendering $\mathbf{R}_\x$ full rank and, consequently, invertible. $\blacksquare$

The $(|\mathscr X|+1)\times(|\mathscr X|+1)$ matrix
$\bar{\mathbf{R}}^\res_\x$ is defined from the rate matrix $\mathbf{R}^\res$ by replacing its line corresponding to vertex $\x$ by $\boldsymbol{\delta}_\emptyset = (1,0\ldots 0)$. The kernel of $\mathbf{R}^\res$ is spanned by $\boldsymbol{\pi}^\res$, which is normalized by $\boldsymbol{\pi}^\res\cdot \boldsymbol{\delta}_\emptyset =1$. Hence, similarly to the above, $\boldsymbol{\delta}_\emptyset$ is not in the coimage of $\mathbf{R}^\res$. This proves that any $|\mathscr X|$ lines of $\mathbf{R}^\res$ and $\boldsymbol{\delta}_\emptyset$ span a $(|\mathscr X|+1)$-dimensional space, so that $\bar{\mathbf{R}}^\res$ has full rank. $\blacksquare$

%------------------------------------------------
\sm{app:linearalgebra}{Algebraic proof of the main result}
%------------------------------------------------

Under the irreducibility assumption, the continuous-time Markov chain generator (rate matrix) \(\mathbf{R}\) has eigenvalue $0$ with multiplicity $1$, allowing us to introduce \(\mathbf{R}_\x\), which is the result of replacing the \(\x\)-th row of the rate matrix by an array of ones. 
Owing to the normalization $\boldsymbol{1}\cdot\boldsymbol{\pi}=1$ of the stationary distribution $\boldsymbol{\pi}$, one has
\begin{equation}
    \label{eq:Rpi}
    \mathbf{R}_\x \, \boldsymbol{\pi} = \boldsymbol{\delta}_\x,
\end{equation}
which can be easily solved since \(\mathbf{R}_\x\) is invertible, in contrast to \(\mathbf{R}\) (see Sec.~\ref{app:invertible}). As put forward in~\cite{aslyamov2023nonequilibrium}, the derivative of the stationary probability in terms of a quantity \(q\) is thus
\begin{equation}\label{eq:deriv}
    \partial_q \boldsymbol{\pi}
    = - \mathbf{R}_\x^{-1}
    \partial_q \mathbf{R}_\x
     \boldsymbol{\pi}.
\end{equation}

We draw attention to the fact that the rate matrix only depends on the input rates in four of its components, both in the positions of $\pm 1$ and in the respective exit rates, see:
\begin{equation}
    \mathbf{R} =\  
    \begin{blockarray}{cccccc}
     & \s(+1) &  & \s(-1) &  \\
    \begin{block}{(ccccc)c}
       &  &  &  &  &  \\
       & -r_{+1}-\ldots &  & r_{-1} &  & \ \s(+1) \\
       &  &  &  &  &  \\
       & r_{+1} &  & -r_{-1}-\ldots &  & \ \s(-1) \\
       &  &  &  &  &  \\
    \end{block}
    \end{blockarray}
\end{equation}
where other rates are left as blank spaces.

Now, we choose \(\x \equiv \s(+1)\), so the input rates in \(\mathbf{R}_\x\) are only present in two elements, as illustrated below:
\begin{equation}\label{eq:rs}
    \mathbf{R}_{\s(+1)} =\  
    \begin{blockarray}{cccccc}
     & \s(+1) &  & \s(-1) &  \\
    \begin{block}{(ccccc)c}
       &  &  &  &  &  \\
       & 1 &  & 1 & & \ \s(+1) \\
       &  &  &  &  &  \\
       & r_{+1} &  & -r_{-1}-\ldots &  & \ \s(-1) \\
       &  &  &  &  &  \\
    \end{block}
    \end{blockarray}
\end{equation}
Applying this choice of $\x$ to Eq.~\eqref{eq:deriv}, the derivative of $\mathbf{R}_{\s(+1)}$ in terms of $r_{\pm 1}$ will have a single nonzero element, leading to the following expressions for the derivatives of the currents:
\begin{equation}\label{eq:currs}
    \partial_{r_{\pm1}} \jj_1 = \pm \left\lbrace 1 - r_{+1} [\mathbf{R}_{\s(+1)}^{-1}]_{\s(+1),\s(-1)} + r_{-1} [\mathbf{R}_{\s(+1)}^{-1}]_{\s(-1),\s(-1)} \right\rbrace \pi_{\s( \pm 1)}
\end{equation}
Notice that Laplace's formula for the determinant provides a suggestive result when evaluated at row $\s(-1)$:
\begin{align}\label{eq:det}
    \mathrm{det} (\mathbf{R}_{\s(+1)}) = &(-1)^{\s(+1)+\s(-1)} r_{+1} \mathrm{det}(\mathbf{R}_{\s(+1),\del (\s(-1), \s(+1))}) - r_{-1} \mathrm{det}(\mathbf{R}_{\s(+1),\del (\s(-1), \s(-1))}) + c \nonumber\\
    =& r_{+1} \mathrm{det} (\mathbf{R}_{\s(+1)}) [\mathbf{R}_{\s(+1)}^{-1}]_{\s(+1),\s(-1)} - r_{-1} \mathrm{det} (\mathbf{R}_{\s(+1)}) [\mathbf{R}_{\s(+1)}^{-1}]_{\s(-1),\s(-1)} +c,
\end{align}
where \(c\) depends on neither input rates, we also used the relation between inverse and minors with notation $\mathbf{R_{\del( \x, \y)}}$ representing the matrix after removal of row $\x$ and column $\y$, and for simplicity the exponent in $(-1)^{\s(+1)+\s(-1)}$ should be interpreted as the sum of the numeric labels given to the respective vertices. Therefore, the derivative of the input current is given by
\begin{equation}\label{eq:dj1}
    \partial_{r_{\pm 1}} \jj_1 = \pm \frac{c \pi_{\s (\pm 1)}}{\mathrm{det} (\mathbf{R}_{\s(+1)})}.
\end{equation}

Now, following the same steps for an arbitrary non-input current $\jj_{e\neq 1}$:
\begin{align}\label{eq:dje}
    \partial_{r_{\pm 1}} \jj_{e \neq 1} = \pm \biggr\lbrace & - r_{+e} 
    (-1)^{\s(+e)+\s(-1)} \mathrm{det} \left( \mathbf{R}_{\s(+1),\del ( 
\s(-1), \s(+e) )} \right) \nonumber\\
    &+ r_{-e} (-1)^{\s(-e)+\s(-1)} \mathrm{det} \left( \mathbf{R}_{\s(+1),\del ( \s(-1), \s(-e) )} \right)
    \biggr\rbrace \frac{\pi_{\s(\pm 1)}}{ \mathrm{det} ( \mathbf{R}_{\s(+1)} )}
\end{align}
and, finally, the ratio of Eqs.~\eqref{eq:dj1} and~\eqref{eq:dje} is
\begin{equation}\label{eq:ratio}
    \frac{\partial_{r_{\pm 1}}\jj_e}{\partial_{r_{\pm 1}}\jj_1} = \frac{
    - r_{+e} (-1)^{\s(+e)+\s(-1)} \mathrm{det} \left( \mathbf{R}_{\s(+1),\del ( \s(-1), \s(+e) )} \right) + r_{-e} (-1)^{\s(-e)+\s(-1)} \mathrm{det} \left( \mathbf{R}_{\s(+1),\del ( \s(-1), \s(-e) )} \right) } {c} \eqqcolon \lambda^1_{e\leftarrow 1}.
\end{equation}

Even more important than this exact expression or how to express $c$, we notice that none of the elements in Eq.~\eqref{eq:ratio} depend on $r_{\pm 1}$, even though Eqs.~\eqref{eq:dj1} and~\eqref{eq:dje} individually present a nonlinear dependence on each of them. This can be seen from the fact that, after removal of row $\s(-1)$ from $\mathbf{R}_{\s(+1)}$, all terms with $r_{\pm 1}$ are either removed or turned into 1 from the involved determinants [see Eq.~\eqref{eq:rs}].

This completes the proof that
\begin{equation}
    \jj_e (\mathbf{r}_1) = \lambda^0_{e\leftarrow 1} + \lambda^1_{e\leftarrow 1}\, \jj_1 (\mathbf{r}_1).\ \blacksquare
\end{equation}

In the case of open CRNs,
the normalization $\boldsymbol{\pi}^\res\cdot\boldsymbol{\delta}_\emptyset=1$
of the stationary distribution $\boldsymbol{\pi}^\res$  imposes that 
the matrix $\bar{\mathbf{R}}^\res_\x$ is invertible
(see Sec.~\ref{app:invertible})
and verifies
\begin{equation}
    \label{eq:Rresxpires}
    \bar{\mathbf{R}}^\res_\x \, \boldsymbol{\pi}^\res = \boldsymbol{\delta}_\x,
\end{equation}
which takes a similar form as in the Markov-chain case [see Eq.~\eqref{eq:Rpi}].
We thus have
$ 
\partial_q \boldsymbol{\pi}^\res
    = - (\bar{\mathbf{R}}^\res_\x)^{-1} \partial_q \bar{\mathbf{R}}^\res_\x
     \boldsymbol{\pi}^\res
$,
similarly to Eq.~\eqref{eq:deriv}.
Also, for \(\x \equiv \s(+1)\), Eq.~\eqref{eq:rs} now becomes
\begin{equation}\label{eq:rsopenCRN}
    \bar{\mathbf{R}}^\res_{\s(+1)} =\  
    \begin{blockarray}{cccccc}
     & \s(+1) &  & \s(-1) &  \\
    \begin{block}{(ccccc)c}
       &  &  &  &  &  \\
       & \bullet &  & \bullet & & \ \s(+1) \\
       &  &  &  &  &  \\
       & r_{+1} &  & -r_{-1}-\ldots &  & \ \s(-1) \\
       &  &  &  &  &  \\
    \end{block}
    \end{blockarray}
\:,
\end{equation}
where $\bullet\in\{0,1\}$ are constants independent of rates.
Then using $\pi^\res_{\s(+f)}=\pi^\res_\emptyset=1 (\forall f\in\mathscr F)$,
we rewrite Eq. (6) of the main text as
\begin{equation}
\label{eq:jres_piempty}
\jj_f = r_{+f}\pi^\res_{\s(+f)} - r_{-f} \pi^\res_{\s(-f)} 
\:.
\end{equation}
Since we also have
$\jj_e = r_{+e}\pi^\res_{\s(+e)} - r_{-e} \pi^\res_{\s(-e)} (\forall e\in\mathscr E)$ 
by definition,
we see that the expression of the stationary current in open CRNs
is linear in $\boldsymbol{\pi}^\res$,
and takes the same form as in the Markov-chain case that we just proved.
The rest of the proof thus follows similar steps as above along Eq.~\eqref{eq:currs} to Eq.~\eqref{eq:ratio} (with the extra species~$\emptyset$). $\blacksquare$

%------------------------------------------------
\sm{app:laplacespace}{Linearity in the Laplace domain}
%------------------------------------------------

\noindent
\textit{Detailed derivation of the result in Laplace domain, Eq. (4) of the main text}
\medskip

\begin{figure}[t]
    \centering
    \includegraphics[width=.4\textwidth]{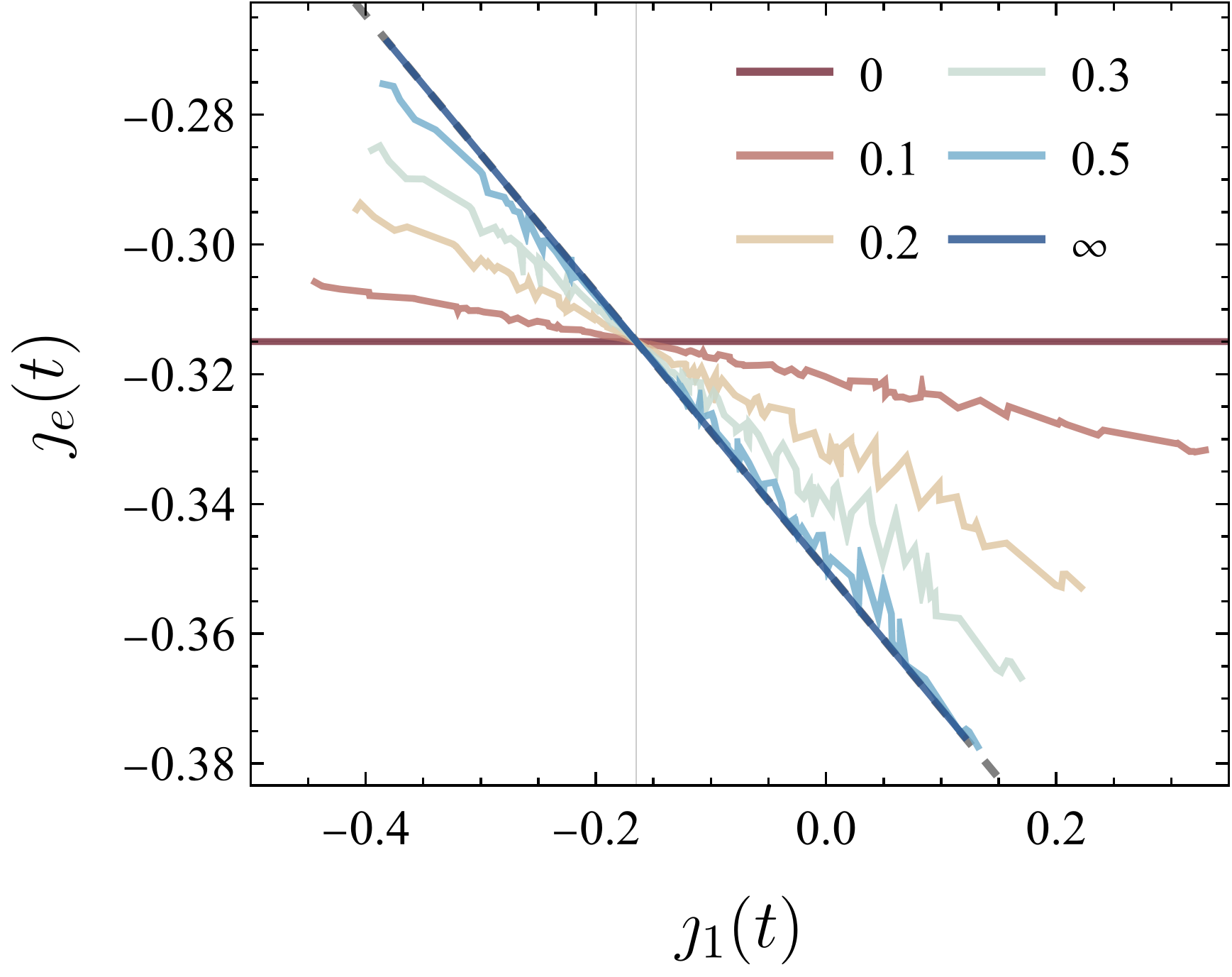}%
    \hspace{15pt}%
    \includegraphics[width=.38\textwidth]{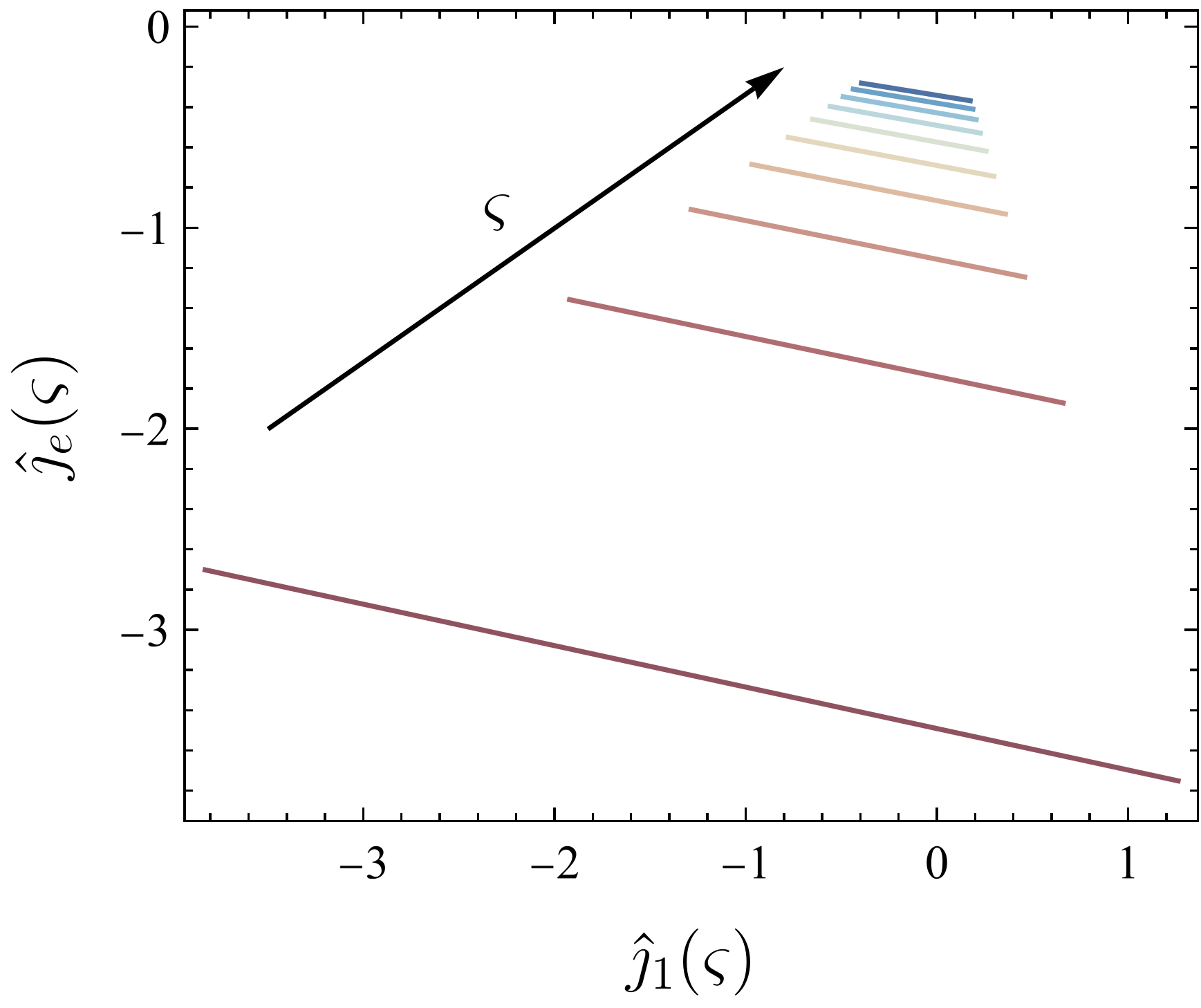}%
    \caption{Time-dependent currents and their Laplace transforms for a process that starts in a stationary distribution and then $\r_1$ gets quenched to a random value. Left panel: Distinct colors represent relaxation times after the quench; for $t=0$, the values of $\jj_1(0)$ change due to the new value of $\r_1$, while $\jj_e(0)$ does not have time to respond to the quench; for long times the currents satisfy the mutual linearity relation for any values of $\r_1$, according to Eq. (2) of the main text, which is represented by a dashed gray line; importantly, the currents are not linearly related for intermediate times. Right panel: The Laplace transformed currents are linearly related for all values of $\si$ as predicted by Eq.~\eqref{eq:result-laplace-app}.}
    \label{fig:finitetime}
\end{figure}

The Laplace transform $\ph(\si)=\int_0^\infty \dd t\: e^{-\si t} \mathbf{p}(t)$ of the time-dependent formal solution of the master equation $\partial_t \mathbf{p}(t) = \mathbf{R}\, \mathbf{p}(t)$ is defined for $\si \in \mathbb C \smallsetminus \operatorname{Sp} \mathbf{R}$ (since $\mathbf{p}(t)$ is a linear combination of terms of the form $t^n e^{\Lambda t} $ for some integers $n\geq 0$ and $\Lambda\in\operatorname{Sp}\mathbf{R}).$ Taking the Laplace transform of the master equation yields $\ph(\si) = (\si \mathbf{1}-\mathbf{R})^{-1} \mathbf{p}(0)$ where $\mathbf{p}(0)$ is the initial condition for $\mathbf{p}(t)$. The resolvent $(\si \mathbf{1}-\mathbf{R})^{-1}$ is defined for $\si \in \mathbb C \smallsetminus \operatorname{Sp} \mathbf{R}$ (from now on we assume that this is the domain of $\si$). From this, we obtain
\begin{equation}\label{eq:dphatdq}
    \partial_q \ph(\si) = (\si \mathbf{1}-\mathbf{R})^{-1} \left( \partial_q \mathbf{R}\right) \ph(\si)
    \:.
\end{equation}
for the derivative in terms of a transition rate $q$.
We now introduce the decomposition $\mathbf{R} = - \mathbf{S}\, \mathbf{V}^\top$ with 
$\mathbf{S}$ and $\mathbf{V}$ two rectangular matrices of dimensions $|\mathscr{X}| \times |\mathscr{E}| $ and elements
\begin{align}
    \label{eq:defS}
    \mathbf{S}_{\x,e} &=  \delta_{\x,\s(-e)} - \delta_{\x,\s(+e)} 
    \\
    \label{eq:defV}
    \mathbf{V}_{\x,e} &=  r_{-e} \, \delta_{\x,\s(-e)} - r_{+e} \, \delta_{\x,\s(+e)} \:.
\end{align}
Here $\mathbf{S}$ is the incidence matrix of the transition graph, and 
$\mathbf{V}$ is a weighted version of it.
This decomposition is used for instance to prove the Markov chain tree theorem (see e.g.\:Chap.\:2 in~\cite{avanzini_methods_2023}).
From the definition in Eq. (1) of the main text for the current, we also have
$\jhat(\si) = - \mathbf{V}^\top \ph(\si)$
(with an obvious notation for the Laplace transform of the current). 
Differentiating $\jhat(\si) $ with respect to a transition rate $q$, and using the Woodbury-type matrix identity
$
\mathbf{1} - \mathbf{V}^\top \big(\si \mathbf{1}+\mathbf{S}\, \mathbf{V}^\top\big)^{-1} \mathbf{S}
= 
\si \, \big(\si \mathbf{1} + \mathbf{V}^\top \mathbf{S}\big)^{-1} 
$,
one obtains from~\eqref{eq:dphatdq}
\begin{equation}\label{eq:djdq}
    \partial_q \jhat(\si) 
    = - \si\, \big(\si \mathbf{1}+ \mathbf{V}^\top\, \mathbf{S} \big)^{-1}\, \left( \partial_q \mathbf{V}^\top \right) \ph(\si)
    \:.
\end{equation}
The Weinstein–Aronszajn identity (known as Sylvester's determinant theorem) implies 
$\det (\mathbf{1}+\si^{-1}\mathbf{V}^\top\, \mathbf{S} )
=
\det (\mathbf{1}+\si^{-1}\mathbf{S}\, \mathbf{V}^\top)
=
\det [(\si \mathbf{1}-\mathbf{R})\si^{-1}]
$,
so that the matrix $\si \mathbf{1}+ \mathbf{V}^\top\, \mathbf{S} $ is invertible
in the considered domain of $\si$.
We can now specialize to variations w.r.t.\:$r_{\pm 1}$.
We first notice from the definition~\eqref{eq:defV} of $\mathbf{V}$ that
\begin{equation}\label{eq:dVdr1}
    \partial_{r_{\pm 1}} \mathbf{V}^\top = \mp | 1\rangle \langle \s(\pm 1)|
    \:,
\end{equation}
where we used bra-ket notation (namely, $| 1\rangle $ is a column vector representing a Kronecker delta for edge $1$ and, similarly, $ \langle \s(\pm 1)|$ is a line vector representing a Kronecker delta for state $ \s(\pm 1)$ ).
We thus obtain from~\eqref{eq:djdq}-\eqref{eq:dVdr1}
\begin{align}
\partial_{r_{\pm 1}} \hat\jj_e (\si) 
&=
\langle e | \partial_{r_{\pm 1}} \jhat(\si) \rangle
=
\pm \si
\,
\langle e | \big(\si \mathbf{1}+ \mathbf{V}^\top\, \mathbf{S} \big)^{-1} | 1\rangle 
\,
\langle \s(\pm 1) | \ph(\si) \rangle
\\
\partial_{r_{\pm 1}} \hat\jj_1 (\si) 
&=
\langle 1 | \partial_{r_{\pm 1}} \jhat(\si) \rangle
=
\pm \si
\,
\langle 1 | \big(\si \mathbf{1}+ \mathbf{V}^\top\, \mathbf{S} \big)^{-1} | 1\rangle 
\,
\langle \s(\pm 1) | \ph(\si) \rangle
\:.
\end{align}
Dividing the first equality by the second and
using the relation between the inverse of a matrix and its minors, we finally arrive at
\begin{equation} \label{eq:djedr1overdj1dr1}
    \frac{\partial_{r_{\pm 1}} \hat\jj_e (\si)}{\partial_{r_{\pm 1}} \hat\jj_1 (\si) }
    =
    (-1)^{1+e}
    \frac
      {
      \det \big(\si \mathbf{1}+ \mathbf{V}^\top\, \mathbf{S} \big)_{ \smallsetminus(1,e)}
      }
      {
      \det \big(\si \mathbf{1}+ \mathbf{V}^\top\, \mathbf{S} \big)_{ \smallsetminus(1,1)}
      }
    \eqqcolon
    \hat \lambda^1_{e\leftarrow 1}(\si)
    \:,
\end{equation}
where $\mathbf{A}_{ \smallsetminus(i,j)}$ is the matrix $\mathbf{A}$ after the removal of row $i$ and column $j$. Since the transition rates are not present in $\mathbf{S}$, matrix $\mathbf{V}^\top\, \mathbf{S}$ depends on the rates $\boldsymbol{r}_1$ only in its row $1$, which means that Eq.~\eqref{eq:djedr1overdj1dr1} is a quantity independent of $\boldsymbol r_{1}$.
With the same argument as in Sec.~\ref{app:linearalgebra}, we thus obtain
\begin{equation}\label{eq:result-laplace-app}
    \hat\jj_e(\r_{1},\si) 
    =
    \hat \lambda^0_{e\leftarrow 1}(\si) 
    +
    \hat \lambda^1_{e\leftarrow 1}(\si) 
    \, 
    \hat \jj_1(\r_{1},\si)
    \:\blacksquare
\end{equation}
This result is illustrated in Fig.~\ref{fig:finitetime}.

\bigskip
\noindent
\textit{Small $\si$ asymptotics and relation to the stationary result, Eq. (2) of the main text}
\medskip

We recall that if $g(t)$ is a function that has a limit $g(+\infty)$ as $t\to +\infty$, then
$g(+\infty)=\lim_{\si\to 0} \si\,\hat g(\si)$,
where $\hat g(\si)$ is the Laplace transform of $g(t)$.
Since the currents converge to their stationary values as $t\to +\infty$, and since $\lambda^0_{e\leftarrow 1}$ is equal to the current $\jj_e$ when $\mathbf{r}_1\to 0$, we have that
\begin{equation}
    \label{eq:limjhats0}
    \lim_{\si\to 0} \,\si\, \hat\jj_e(\r_{1},\si) = \jj_e(\r_1)
\:,
\quad
    \lim_{\si\to 0} \,\si\, \hat \lambda^0_{e\leftarrow 1}(\si)  = \lambda^0_{e\leftarrow 1}
\:,
\quad
    \lim_{\si\to 0} \,\si\, \hat\jj_1(\r_{1},\si) = \jj_1(\r_1)
\:,
\end{equation}
where $\jj_e(\r_1)$, $\lambda^0_{e\leftarrow 1}$, $\jj_1(\r_1)$ are the stationary quantities involved in the mutual linearity of Eq. (2) of the main text.
From~\eqref{eq:result-laplace-app} and Eq. (2) of the main text, this implies  necessarily that $\hat \lambda^1_{e\leftarrow 1}(\si) $ is bounded for $\si\to 0$,
and
$\lim_{\si\to 0}\, [ \si\, 
    \hat \lambda^1_{e\leftarrow 1}(\si) 
    \, 
    \hat \jj_1(\r_{1},\si)
] = \lambda^1_{e\leftarrow 1} \, \jj_1(\r_{1})$
which in turn yields (notice the difference with Eq.~\eqref{eq:limjhats0}):
\begin{equation}
    \label{eq:lambda1lim}
   \lambda^1_{e\leftarrow 1} = \lim_{\si\to 0}  \hat \lambda^1_{e\leftarrow 1}(\si) 
   \:.
\end{equation}

This shows that the mutual linearity in the Laplace domain yields, in the $\si\to 0$ limit, the corresponding stationary relation.
Furthermore, one obtains from~\eqref{eq:djedr1overdj1dr1} another expression of the susceptibility coefficient:
\begin{equation}
    \label{eq:lambda1limratiodet}
   \lambda^1_{e\leftarrow 1} 
   =
   \lim_{\si\to 0} 
      \;
       (-1)^{1+e}
    \frac
      {
      \det \big(\si \mathbf{1}+ \mathbf{V}^\top\, \mathbf{S} \big)_{ \smallsetminus(1,e)}
      }
      {
      \det \big(\si \mathbf{1}+ \mathbf{V}^\top\, \mathbf{S} \big)_{ \smallsetminus(1,1)}
      }
      \:.
\end{equation}
Notice that the limit is not taken trivially since numerator and denominator both go to 0 as $\si\to 0$ (see App.~\ref{app:graphrepresentation}). Intuitively, Eq.~\eqref{eq:lambda1lim} means that the slope of lines in the right panel of Fig.~\ref{fig:finitetime} are converging to $\lambda^1_{e\leftarrow 1}$ for small values of $\si$.

%------------------------------------------------
\sm{app:graphrepresentation}{Spanning-tree ensemble for the current-current susceptibility}
%------------------------------------------------

\noindent
\textit{Relation between the $\si\to 0$ asymptotics and the spanning-tree representations: the denominator of Eq.~\eqref{eq:lambda1limratiodet}}
\medskip

Let us now analyze the limit in Eq.~\eqref{eq:lambda1limratiodet}, starting by its denominator, which acts as a normalization.
We first remark that 
$
(\mathbf{V}^\top\, \mathbf{S} )_{ \smallsetminus(1,1)}
=
(\mathbf{V}_{ \smallsetminus(\cdot,1)})^\top\, \mathbf{S} _{ \smallsetminus(\cdot,1)}
$
where $ _{ \smallsetminus(\cdot,1)}$  indicates the removal of column $1$. Then, using a Weinstein–Aronszajn identity,
we obtain that the denominator rewrites as
\begin{equation}
    \label{eq:denom1}
    \det \big(\si \mathbf{1}+ \mathbf{V}^\top\, \mathbf{S} \big)_{ \smallsetminus(1,1)}
    =
    \si^{E-X-1} \det \big( \si \mathbf{1} - \mathbf{R}_{\smallsetminus 1} \big)
\end{equation}
where
$
\mathbf{R}_{\smallsetminus 1} 
= 
-\mathbf{S} _{ \smallsetminus(\cdot,1)}
\,
(\mathbf{V}_{ \smallsetminus(\cdot,1)})^\top
$
represents the rate matrix of the Markov chain deprived of edge $1$.
For simplicity, we denote by $X=|\mathscr X|$ and $E=|\mathscr E|$ the numbers of states and edges respectively.
In practice, $\mathbf{R}_{\setminus 1}$ is obtained from $\mathbf{R}$ by taking the limits $r_{\pm 1}\to 0$, and since we have assumed that edge $1$ is not a bridge, it still represents an irreducible Markov chain on a network (so that $0$ is an eigenvalue of $\mathbf{R}_{\setminus 1}$ with multiplicity $1$).
This implies that Eq.~\eqref{eq:denom1} behaves as $\si^{E-X}$ as $\si\to 0$.
To analyze the prefactor of this small-$\si$ asymptotics,
we use Jacobi's formula, 
$
\det\,(\mathbf{A}+\varepsilon \mathbf{B}) 
=
\det\, \mathbf{A} + \varepsilon \operatorname{tr} 
 \big[ 
 (\operatorname{adj} \mathbf{A}) \mathbf{B}
 \big]
 + O(\varepsilon^2)
$,
where $\operatorname{adj} \mathbf{A}$ denotes the adjugate matrix of $\mathbf{A}$.
Since 
$
\mathbf{R}_{\smallsetminus 1}
$
is stochastic, we have 
$
\det \mathbf{R}_{\smallsetminus 1} = 0
$
and thus
\begin{equation}
    \label{eq:denom2}
    \det \big(\si \mathbf{1}+ \mathbf{V}^\top\, \mathbf{S} \big)_{ \smallsetminus(1,1)}
    \ \
    \underset{\si\to 0}{\sim}
    \ \
    \si^{E-X} 
    \operatorname{tr} \big( \operatorname{adj} (-\mathbf{R}_{\smallsetminus 1}) \big)
    \:.
\end{equation}
In this expression, the trace of the adjugate of $-\mathbf{R}_{\smallsetminus 1} $
has a clear graph-theoretical interpretation (see e.g.~Chap.~2 in~\cite{avanzini_methods_2023}) in terms of rooted spanning trees of the generator
$\mathbf{R}_{\smallsetminus 1}$,
which are the spanning trees of the original process that do not contain edge $1$, as we now explain. We recall that in a directed graph, a spanning tree $\mathscr{T}_\x$ with root $\x$ is a subset of transitions such that every vertex of the graph is connected to $\x$ via a unique path and every transition along such path is pointing toward $\x$. Let $w (\mathscr T_\x)$ be the product of the transition rates in $\mathscr{T}_\x$. 
Then, the trace of the adjugate of $-\mathbf{R}_{\smallsetminus 1} $ represents the sum of $w (\mathscr T_\x)$ over all the possible rooted spanning trees of $\mathbf{R}_{\smallsetminus 1} $~\cite{avanzini_methods_2023}.
This gives the final expression of the denominator of the susceptibility in Eq.~\eqref{eq:lambda1limratiodet}, in the small-$\si$ asymptotics, as a sum over rooted spanning trees:
\begin{equation}
    \label{eq:denom2.2}
    \det \big(\si \mathbf{1}+ \mathbf{V}^\top\, \mathbf{S} \big)_{ \smallsetminus(1,1)}
    \ \
    \underset{\si\to 0}{\sim}
    \ \ 
    \si^{E-X} 
     \tau_{\del 1}
    \:,
\end{equation}
where $\tau_{\setminus 1} = \sum_{\x}
   \sum_{\mathscr T_\x \subseteq \mathscr{G} \;, 1 \not\subseteq \mathscr T_\x}
   w\big(\mathscr T_\x \big)$ indicates the spanning tree polynomial, and the sum runs over all rooted spanning trees of the original graph which  do not contain edge $1$.

\bigskip
\bigskip
\noindent
\textit{Relation between the $\si\to 0$ asymptotics and the graph-theoretic representations: the numerator of Eq.~\eqref{eq:lambda1limratiodet}}
\medskip

The numerator is less simple to depict directly.
To understand it, we introduce a matrix $
(\mathbf{V}^\top\, \mathbf{S} )^\prime
$
obtained by replacing row $1$ of $\mathbf{V}^\top\, \mathbf{S}$ by an array with $-1$ in position $e$ and $0$ elsewhere. Then, we compute the determinant by using Laplace's expansion of the modified row:
\begin{equation}
     \label{eq:keylaplaceexpansion}
    \det
    \big[
    \si \mathbf{1} + (\mathbf{V}^\top\, \mathbf{S} )^\prime 
    \big]
    =
    \si \det 
    \big(
      \si \mathbf{1}+ \mathbf{V}^\top\, \mathbf{S}
    \big)_{ \smallsetminus(1,1)}
    \ 
    -
    \
    (-1)^{1+e}
      \det \big(\si \mathbf{1}+ \mathbf{V}^\top\, \mathbf{S} \big)_{ \smallsetminus(1,e)}
      \; .
\end{equation}
We observe from Eq.~(\ref{eq:lambda1limratiodet}) that the first term on the rhs of Eq.~(\ref{eq:keylaplaceexpansion}) does not contribute in the limit $\si \to 0$.  This means that
\begin{equation}
    \label{eq:lambda1limratiodetBISprime}
   \lambda^1_{e\leftarrow 1} 
   =
   -\lim_{\si\to 0} 
    \frac
      {
    \det
    \big[
    \si \mathbf{1} + (\mathbf{V}^\top\, \mathbf{S} )^\prime 
    \big]
    \quad\:
      }
      {
      \det \big(\si \mathbf{1}+ \mathbf{V}^\top\, \mathbf{S} \big)_{ \smallsetminus(1,1)}
      }
      \:.
\end{equation}

Given the asymptotics of Eq.~(\ref{eq:denom2.2}), we need to understand the behavior  $O(\si^{E-X})$ of 
the numerator of Eq.~\eqref{eq:lambda1limratiodetBISprime}.
To do so, the key point is to rewrite 
$(\mathbf{V}^\top\, \mathbf{S} )^\prime$
as a product ${\mathbf{V}^\star}^\top\, {\mathbf{S}^\star}$
that, using a Weinstein–Aronszajn identity, will allow coming back to the space of states instead of edges.
For this purpose, we introduce a new state $\star$ (placed before the other states in matricial representations), 
and define a 
$(X+1)\times E$ 
matrix ${\mathbf{S}^\star}$ 
obtained from
$\mathbf{S}$ 
by adding a row on top (corresponding to state $\star$)
with $-1$ in position $e$, and 0 elsewhere.
Complementarily, we define a 
$(X+1)\times E $ 
matrix ${\mathbf{V}^\star}$ 
obtained by adding a line on top with $1$ in position $1$, and 0 elsewhere and by replacing 
rates $r_{\pm 1}$ with 0. This last modification only affects column $1$ of $\mathbf{V}$ and leaves $\lambda^1_{e\leftarrow 1}$ unchanged [see Eq.~(\ref{eq:lambda1limratiodet})] because $\lambda^1_{e\leftarrow 1}$ does not depend on $r_{\pm 1}$.
Then, it's a simple matter of computation to check that 
$
(\mathbf{V}^\top\, \mathbf{S} )^\prime
=
{\mathbf{V}^\star}^\top\, {\mathbf{S}^\star}
$.

Such matrices ${\mathbf{S}^\star}$ and ${\mathbf{V}^\star}^\top$ do not represent (weighted) incidence matrices, but we can still use a Weinstein–Aronszajn identity to write
\begin{equation}
    \label{eq:towardsWstarwoodbury}
    \det
    \big[
    \si \mathbf{1} + (\mathbf{V}^\top\, \mathbf{S} )^\prime 
    \big]
=
   \si^{E-X-1}
      \det
    \big(
    \si \mathbf{1} + {\mathbf{S}^\star} {\mathbf{V}^\star}^\top
    \big)
    \:.
\end{equation}

Our goal now is to represent this last determinant as a sum over spanning trees (in the small $\si$ asymptotics).
The $(X+1)\times(X+1)$  matrix $-{\mathbf{S}^\star} {\mathbf{V}^\star}^\top$
does not represent the rate matrix of a stochastic process. However, 
its down-right core $-({\mathbf{S}^\star} {\mathbf{V}^\star}^\top)_{\setminus(\star,\star)}$
is equal to $\mathbf R_{\setminus 1}$, 
which we met previously and is the generator $\mathbf R$ where rates $r_{\pm 1}$ are set to $0$,
i.e.~the generator for the transition graph where edge $1$ is removed.
Then, for the remaining of ${\mathbf{S}^\star} {\mathbf{V}^\star}^\top$:
\begin{itemize}
    \item The first column, corresponding to exiting state $\star$, has entries $-1$ in $\s(+1)$, $1$ in $\s(-1)$ and $0$ elsewhere\footnote{For simplicity the entries of such $(X+1)\times(X+1)$ matrices are labelled as $(\star,1,...,X)$.}. 
    \item The first row, corresponding to entering state $\star$, has entries $r_{+e}$ in $\s(+e)$, $-r_{-e}$ in $\s(-e)$ and $0$ elsewhere. 
\end{itemize}
To express~\eqref{eq:towardsWstarwoodbury} as a sum over spanning trees, we use the multilinearity of the determinant along the first line
of the matrix $ \si \mathbf{1} + {\mathbf{S}^\star} {\mathbf{V}^\star}^\top$
to write
\begin{align}
\label{eq:det1SstarVstarT}
\det
\big(
 \si \mathbf{1} + {\mathbf{S}^\star} {\mathbf{V}^\star}^\top
\big)
=
\si
\det 
\big(
\si \mathbf{1} - \mathbf{R}_{\setminus 1}
\big)
-r_{+e}
\det 
\big[
\si (\mathbf{1} -|\star\rangle\langle\star|) - \mathbf{R}^\star_{\s (+e)}
\big]
+r_{-e}
\det 
\big[
\si (\mathbf{1} -|\star\rangle\langle\star|) - \mathbf{R}^\star_{\s (-e)}
\big]
\end{align}
The operators $ \mathbf{R}^\star_{\s (\pm e)}$ are $(X+1)\times(X+1)$ matrices
that preserve probability (i.e.\: the constant vector is a left null vector) and are defined from $(-{\mathbf{S}^\star} {\mathbf{V}^\star}^\top) $ where:
\begin{itemize}

\item column $\s(+e)$ of  $ \mathbf{R}^\star_{\s (+e)}$, corresponding to exiting state $\s(+e)$, is replaced with a $+1$ on the first line (corresponding to vertex~$\star$), $-1$ on the diagonal and $0$ elsewhere:
\begin{equation}\label{eq:manipulationstochasticity1}
\text{column $\s (+e)$ in  $ \mathbf{R}^\star_{\s (+e)}$}: 
  \begin{pmatrix}
    1 \\ \vdots \\ 0 \\ \vdots \\ -1 \\ \vdots \\ 0
  \end{pmatrix}~
  \begin{array}{@{} l @{}}
    \rightarrow \star \\ \\ \\ \\ \rightarrow \s (+e) \\  \\ \mathstrut
  \end{array}
\end{equation}
\item column $\s(-e)$ of  $ \mathbf{R}^\star_{\s (-e)}$, corresponding to exiting state $\s(-e)$, is replaced with a $+1$ on the first line (corresponding to vertex $\star$), $-1$ on the diagonal and $0$ elsewhere:
\begin{equation}\label{eq:manipulationstochasticity2}
\text{column $\s (-e)$ in  $ \mathbf{R}^\star_{\s (-e)}$}: 
  \begin{pmatrix}
     1 \\ \vdots \\ 0 \\ \vdots \\ -1 \\ \vdots \\ 0
  \end{pmatrix}~
  \begin{array}{@{} l @{}}
    \rightarrow \star  \\ \\ \\ \\ \rightarrow \s (-e) \\  \\   \mathstrut
  \end{array}
\end{equation}
\item The remaining entries of the first row in  $ \mathbf{R}^\star_{\s (\pm e)}$ are $0$.
\end{itemize}

As before, the remaining entries of the $X\times X$ bottom-right block of these matrices 
are those of $\mathbf{R}_{\setminus 1}$.
In these manipulations, we used the fact that, in Eq.~\eqref{eq:det1SstarVstarT},
the determinant involving $\mathbf{R}^\star_{\s(\pm e)}$ does not depend on 
the content of column $\s(\pm e)$ (beyond its first element),
as seen by a Laplace expansion of the determinant along the first line. As a consequence, we are free to fix the content of column  $\s(\pm e)$ in  $\mathbf{R}^\star_{\s(\pm e)}$. The choice in Eqs.~(\ref{eq:manipulationstochasticity1})-(\ref{eq:manipulationstochasticity2}) ensures stochasticity.

Operators $\mathbf{R}^\star_{\s(\pm e)}$ are interpreted as follows:
a state $\star$ is connected 
to $\s(+1)$ with weight $1$,
to $\s(-1)$ with weight $-1$,
and from $\s(+e)$ (resp.~from $\s(-e)$) with weight $1$.
These transitions are unidirectional.
Furthermore, the only outgoing transition from $\s(+e)$ (resp.~from $\s(-e)$) is to state $\star$ [in compliance with Eqs.~(\ref{eq:manipulationstochasticity1})-(\ref{eq:manipulationstochasticity2})].
Graphically:
\begin{center}\label{fig:Rstar}
    \includegraphics[width=.5\columnwidth]{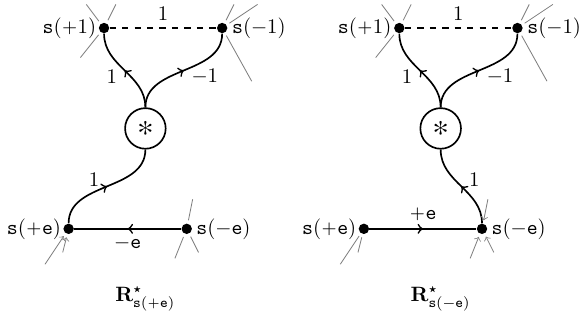}
\end{center}
where all the arrowed edges are strictly unidirectional.

Notice from Eqs.~\eqref{eq:lambda1limratiodetBISprime}-\eqref{eq:towardsWstarwoodbury}
that, to compute $\lambda^1_{e\leftarrow 1}$,
we are interested in the behavior of order $O(\si)$ of Eq.~\eqref{eq:det1SstarVstarT}.
The first term in the r.h.s.~of~\eqref{eq:det1SstarVstarT} is 
$
\si^2 
\operatorname{tr}\,
\operatorname{adj}\,
(-\mathbf{R}_{\setminus 1})
$
and does not contribute.
To understand the two remaining terms of this equation,
we use the following property:
if $\mathbf{R}^\star$ is a $(X+1)\times (X+1)$ matrix
that preserve probability (with the first line corresponding to state $\star$),
we have:
\begin{align}
\det
\big[
\si (\mathbf{1} -|\star\rangle\langle\star|) - \mathbf{R}^\star
\big]
&
\underset{\si\to 0}{=}
\det
\big(
 - \mathbf{R}^\star
\big)
+
\si
\operatorname{tr}
\big[
\big(\mathbf{1} -|\star\rangle\langle\star|\big) 
\operatorname{adj}\,
(- \mathbf{R}^\star)
\big]
\nonumber
\\
&=
\si
\operatorname{tr}
\,
\operatorname{adj}\,
(- \mathbf{R}^\star)
-
\si
\operatorname{tr}
\big[
\,
|\star\rangle\langle\star|
\,
\operatorname{adj}\,
(- \mathbf{R}^\star)
\big]
\nonumber
\\
\label{eq:detnostar}
&=
\si
\sum_{\x \neq \star}
\big(
\operatorname{adj}\,
(- \mathbf{R}^\star)
\big)_{\x \x} 
\:.
\end{align}
All in all, applying this to Eq.~\eqref{eq:det1SstarVstarT} gives:
\begin{equation}\label{eq:detSVstarfirstorder}
\det
\big(
 \si \mathbf{1} + {\mathbf{S}^\star} {\mathbf{V}^\star}^\top
\big) \underset{\si\to 0}{=}  \si  \left[ - r_{+e} \sum_{\substack{\x\neq \star \\ \x\neq \s(+e)} }
\big( 
\operatorname{adj}\,
(-\mathbf{R}^\star_{\s (+e)})
\big)_{\x\x} 
+ r_{-e}  \sum_{\substack{\x\neq \star \\ \x\neq \s(-e)} }
\big( 
\operatorname{adj}\,
(-\mathbf{R}^\star_{\s (-e)}) 
\big)_{\x\x} \right] 
\:.
\end{equation}
In this expression, the sums represent a sum over the spanning trees of $\mathbf{R}^\star_{\s(\pm e)}$
rooted in every state except $\star$ and $\s(\pm e)$. Here we used that $\big(\operatorname{adj}\,(-\mathbf{R}^\star_{\s (\pm e)})\big)_{\s(\pm e), \s(\pm e)} \; = 0$. Indeed, column $\s(+e)$ (resp.~$\s(-e)$) is linearly independent of the remaining columns of $\mathbf{R}^\star_{\s( +e)}$ (resp.~$\mathbf{R}^\star_{\s(- e)}$) as can be seen from Eqs.~(\ref{eq:manipulationstochasticity1})-(\ref{eq:manipulationstochasticity2}), which implies that the rank of the matrices
$(\mathbf{R}^\star_{\s (\pm e)})_{\setminus (\s(\pm e), \s(\pm e))}$
is not maximal\footnote{
We have:
$\rank
(\mathbf{R}^\star_{\s (\pm e)})_{\setminus (\s(\pm e), \s(\pm e))}
\leq
\rank
(\mathbf{R}^\star_{\s (\pm e)})_{\setminus (\cdot, \s(\pm e))}
=
\rank
(\mathbf{R}^\star_{\s (\pm e)}) -1
\leq
X-1
$
since
$
\rank
(\mathbf{R}^\star_{\s (\pm e)}) \leq X
$
(this matrix is stochastic).
}.

A few remarks follow. Since the only outgoing transition from $\s (+e)$ (resp.~$\s(-e)$) is to state $\star$,  $\mathbf{R}^\star_{\s( +e)}$ (resp.~$\mathbf{R}^\star_{\s(- e)}$) does not depend on $r_{+e}$ (resp.~$r_{-e}$).
This is consistent with the fact that the numerator in Eq.~(\ref{eq:lambda1limratiodet}) is a linear function of every rate $r_{\pm e}$ 
(as seen from the multilinearity of the determinant and the fact that $\si \mathbf{1}+ \mathbf{V}^\top\, \mathbf{S}$ depends on $r_{\pm e}$ only through column~$e$). 
Notice also that the ensembles of spanning trees of the two operators $\mathbf{R}^\star_{\s (\pm e)}$ are different.
Nevertheless, a bijection exists between the subset of spanning trees in $\mathbf{R}^\star_{\s (+e)}$ containing the transition $-e$ and the subset of spanning trees in $\mathbf{R}^\star_{\s (- e)}$ containing the transition $+e$. This comes from the definition of spanning tree, which states that every vertex has at most one outgoing transition (zero if it is the root). Consequently, all terms containing products $r_{\pm e} r_{\mp e}$ cancel in the summation of Eq.~(\ref{eq:detSVstarfirstorder}). 

Finally,  we go further and eliminate state $\star$.
Notice that in~\eqref{eq:detSVstarfirstorder}, for $\mathbf{R}^\star_{\s(\pm e)}$,
every spanning tree must pass through state $\star$ either by containing
the path from $\s(\pm e)$ to $\s(+1)$ with weight $1 $
or the path from $\s(\pm e)$ to $\s(-1)$ with weight $-1$. This owns up to the fact that neither  $\s(\pm e)$ nor $\star$ are the root of the tree. 
In other words, 
\begin{eqnarray}\label{eq:Rcontract}
\sum_{\substack{\x\neq \star \\ \x\neq \s(\pm e)} }
\big(
\operatorname{adj}\,
(- \mathbf{R}^\star_{\s(\pm e)})
\big)_{\x\x}
\ 
&=&
\ 
\sum_{ \x\neq \s(\pm e) }
\big(
\operatorname{adj}\,
(- \mathbf{R}_{\setminus 1}^{\s(\pm e)\to\s(+1)})
\big)_{\x\x}
\ 
-
\ 
\sum_{ \x\neq \s(\pm e) }
\big(
\operatorname{adj}\,
(- \mathbf{R}_{\setminus 1}^{\s(\pm e)\to\s(-1)})
\big)_{\x\x} \label{eq:deletion-contractionexpression} 
\:.
 \end{eqnarray}
Here
$
\mathbf{R}_{\setminus 1}^{\s(\pm e)\to\s(+1)}    
$
are $X\times X$ rate matrices (with positive rates)
built from $\mathbf{R}_{\setminus 1}$ by:
(\textit{i}) removing every outgoing transition from $\s(\pm e)$, and
(\textit{ii}) 
    adding one unidirectional edge,
    from $\s(\pm e)$ to $\s(+1)$, with rate $1$. 
The matrices
$
\mathbf{R}_{\setminus 1}^{\s(\pm e)\to\s(-1)}    
$
are defined in a similar manner.

\begin{figure}
    \centering
    \includegraphics[width=.45\columnwidth]{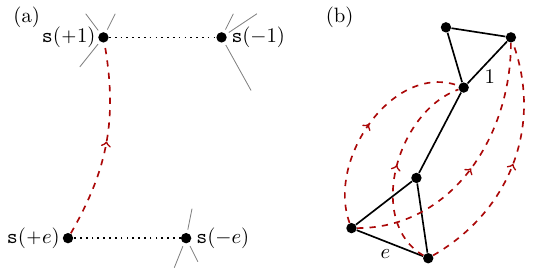}
    \caption{%
    (a) Illustration of the term $\tau_{\del 1, e}^{\con \s(+e) \to \s(+1)}$ entering in the numerator of Eq.~(\ref{eq:lambda1_thirdexpression}), which is Eq. (5) of the main text. In order to compute the susceptibility of an output current $e$ with respect to an input current $1$, one builds a spanning-tree ensemble. Starting from the original network $\mathscr{G}$, one first removes edge $1$ and edge $e$ (in dotted black), and then connects input and output vertices by adding a directed edge from $\s(+e)$ to $\s(+1)$ with unit rate (in dashed red).
    (b)~Example network ($\mathscr{G}$, in black) with two cycles connected by a bridge.
    The spanning trees involved in the expression of the susceptibility Eq.~(\ref{eq:lambda1_thirdexpression}) involve one of the
    additional edges indicated in dashed red. Such red edge is contracted while $1$ and $e$ are deleted.}
    \label{fig:2}
\end{figure}

We now explicit the connection to graph theory and employ notations of the deletion-contraction paradigm to express the spanning tree polynomials. 
We already pointed out that for any stochastic operator $\mathbf R$, $\operatorname{tr} \operatorname{adj} (-\mathbf{R})$ represents the sum of the product of rates $w (\mathscr T_{\x})$ of the rooted spanning trees $ \mathscr T_{\x}$ of the graph $\mathscr G$ associated to $\mathbf R$:
\begin{equation}
\operatorname{tr} \left(\operatorname{adj} (-\mathbf{R})\right)
=
    \sum_{\x} \sum_{\mathscr T_{\x} \subseteq \mathscr G } w (\mathscr T_{\x})
    \:.
\end{equation}
The key observation in Eq.~(\ref{eq:Rcontract}) is that every tree spanning
for $\mathbf{R}_{\setminus 1}^{\s(\pm e)\to\s(+1)}$
has its root different from $\s(\pm e)$ and must contain the transition $\s(\pm e)\to\s(+1)$ (with rate $1$).
Likewise, every tree spanning for $\mathbf{R}_{\setminus 1}^{\s(\pm e)\to\s(-1)}$ has its root different from $\s(\pm e)$ and must contain the transition  $\s(\pm e)\to\s(-1)$ (with rate $1$). 
Then, the terms on the r.h.s of Eq.~(\ref{eq:Rcontract}) can be re-expressed as spanning-tree polynomials of a modified graph where $(i)$ edge $1$ and $e$ are deleted and $(ii)$ unidirectional edge $ \s(\pm e)\to\s(\pm 1)$ is added and contracted, see Fig.~\ref{fig:2} for an illustration of this procedure.
If we denote such modified graphs by $\mathscr G_{\del 1 , e}^{\s(\pm e) \to \s(\pm 1)}$, the full susceptibility is finally expressed as follows (using Eqs.~\eqref{eq:denom2.2}, \eqref{eq:detSVstarfirstorder}-\eqref{eq:deletion-contractionexpression}  in Eq.~\eqref{eq:lambda1limratiodetBISprime}):
\begin{align}
\label{eq:lambda1_thirdexpression}
    \lambda^1_{e\leftarrow 1}
    =  \displaystyle
    r_{+e}\,
    \frac
    {
    \tau_{\del 1,\, e}^{ \con \s(+e) \to \s(+1)}-\tau_{\del 1,\, e}^{ \con \s(+e)\to \s(-1)}}{
    \tau_{\setminus 1}
    }
    - r_{-e}\,
    \frac{
    \tau_{\setminus 1,\, e}^{ \con \s(-e)\to \s(+1)}-\tau_{\setminus 1,\, e}^{\con \s(-e)\to \s(-1)}
    }
    {
    \tau_{\setminus 1}
    }
\end{align}
with
\begin{equation}\label{eq:taudefinition}
    \tau_{\del 1, e}^{\con \s(\pm e) \to \s(+1)} = \sum_{\x} \sum_{\substack{\mathscr T_{\x} \subseteq \mathscr G_{\del 1, e}^{\s(\pm e) \to \s(+1)} \\ \s(\pm e) \to \s(+1) \subseteq \mathscr T_{\x}}} w (\mathscr T_{\x}),
\end{equation}
and similarly for $\s(-1)$.

A few remarks follow. First, contracting edge  $ \s(\pm e)\to\s(\pm 1)$ in (\ref{eq:lambda1_thirdexpression})-(\ref{eq:taudefinition}) accounts for the removal of outgoing transitions from $s(\pm e)$ in $\mathbf{R}_{\setminus 1}^{\s(\pm e)\to\s(\pm1)}$ because every vertex in a rooted tree has at most one outgoing transition.
Secondly, we removed edge~$e$ from the spanning tree polynomials in (\ref{eq:lambda1_thirdexpression})-(\ref{eq:taudefinition}) thanks to the compensation between quadratic terms $ \propto r_{\pm e} r_{\mp e}$ discussed above. Thus operatively, the spanning tree polynomials entering Eq.~(\ref{eq:lambda1_thirdexpression}) are obtained directly from the original graph by $(i)$ removing the input and output edges and $(ii)$ adding and contracting a unidirectional edge which connects directly vertices $\s(\pm e)$ to vertices $\s(\pm 1)$. 
Graphically:
\begin{center}
\includegraphics[width=.9\columnwidth]{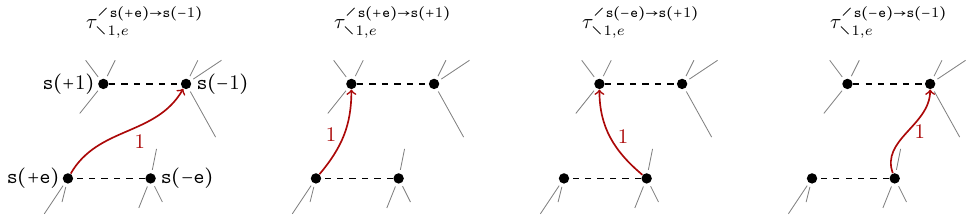}
\end{center}
where the dashed edges are deleted and the red edges are unidirectional and contracted with rate $1$. Remark that if an edge $ \s(\pm e)\to\s(\pm 1)$ is already present in the original graph, it is simply replaced by the unidirectional red edge depicted above.

%------------------------------------------------
\sm{app:bridge}{Why did the perturbation cross the bridge?}
%------------------------------------------------

If the removal of an edge separates the network into two disconnected networks, this edge is called a bridge $\mathcal{B}$, and we will refer to these two subnetworks as islands. For what is concerned here, the bridge can also form a subnetwork, possibly with cycles, provided it shares a single vertex with each island. Consider that the perturbed edge belongs to the first island $\mathcal{I}_1$; in general, all probabilities of the vertices in $\mathcal{I}_1$ will change with the perturbation, which includes the vertex that connects to $\mathcal{B}$. Since the stationary current over a bridge is zero, the probability of the vertex on its opposite end will compensate for the said change, and thus all probabilities of vertices in $\mathcal{I}_2$ will change.

Notice that each spanning tree $\mathscr{T}$ can be split into three parts: $\mathscr{T} = \mathscr{T}^{\mathcal{I}_1} \cup \mathscr{T}^{\mathcal{B}} \cup \mathscr{T}^{\mathcal{I}_2}$, highlighting the island/bridge to which the branches belong. From the Markov chain tree theorem, the probability of a vertex $\x$ belonging to $\mathcal{I}_2$ is
\begin{equation}
    \pi_\x \propto \sum_{\mathscr{T}_\x}  w(T) =
    \left( \sum_{\mathscr{T}_\mathcal{B}^{\mathcal{I}_1}}  w(\mathscr{T}_\mathcal{B}^{\mathcal{I}_1}) \right)
    \left( \sum_{\mathscr{T}_{\mathcal{I}_2}^{\mathcal{B}}}  w(\mathscr{T}_{\mathcal{I}_2}^{\mathcal{B}}) \right)
    \left( \sum_{\mathscr{T}_\x^{\mathcal{I}_2}}  w(\mathscr{T}_\x^{\mathcal{I}_2}) \right).
\end{equation}
where $\mathscr{T}_\mathcal{B}^{\mathcal{I}_1}$ are all spanning trees of island 1 rooted at the vertex shared with the bridge, and $\mathscr{T}_{\mathcal{I}_2}^{\mathcal{B}}$ are the spanning trees of the bridge rooted at the vertex shared with island 2. When the bridge is a single edge, the latter term is simply its rate directed to the vertex shared with $\mathcal{I}_2$.

The ratio between the probabilities of two vertices $\x, \y \in \mathcal{I}_2$ will then only depend on rates from island 2, which are not affected by the perturbations, and is thus constant. Consequently, when rates are perturbed in island 1, \textit{(i)} all probabilities in island 2 will change by the same multiplicative factor; \textit{(ii)} all currents will change by the same multiplicative factor; \textit{(iii)} if $\mathcal{I}_2$ satisfies detailed balance, no perturbation in $\mathcal{I}_1$ will make currents flow in $\mathcal{I}_2$, thus their susceptibilities are zero.

Since the ratio of two currents in island 2 is fixed, $\jj_e (\r_i) / \jj_{e'}(\r_i) = \jj_e (\r_i^*) / \jj_{e'}(\r_i^*)$ for $\{e,e'\} \in \mathcal{I}_2$ (if $\jj_{e'}$ is not strictly zero, which happens in detailed balance conditions), their affine and linear coefficients with respect to the input edge satisfy
\begin{equation}
    \det \begin{pmatrix} \lambda_{e \leftarrow i}^0 & \lambda_{e \leftarrow i}^1 \\ \lambda_{e' \leftarrow i}^0 & \lambda_{e' \leftarrow i}^1 \end{pmatrix} = 0,
\end{equation}
which, using Eq. (3) of the main text,
implies $\jmath_{e'}(\mathbf{r}_i) = ( \lambda^1_{e'\leftarrow i}/ \lambda^1_{e\leftarrow i} )\jmath_{e}(\mathbf{r}_i)$,
i.e.~all currents are strictly linear one to another (without affine coefficient) when they live on a different island than the input edge, and they are controlled by this input edge.
Of course this does not mean at all that $\lambda_{e' \leftarrow e}^0 = 0$.

%------------------------------------------------
\sm{app:mutual-positivity}{Symmetries of the the susceptibility when reversibility holds}
%------------------------------------------------

Let us assume that the dynamics is reversible, namely, that detailed balance holds with respect to some equilibrium distribution $\boldsymbol{\pi}^\eq$:
\begin{equation}
    \label{eq:defDB}
    \pi^\eq_{\s(+e)} r_{+e} 
    =  
    \pi^\eq_{\s(-e)} r_{-e} 
    \quad 
    \forall{e\in\mathscr E}
    \:.
\end{equation}
This implies (see Eq. (1) of the main text) that the stationary currents are $0$.
Removing any edge $i$ by setting $r_{\pm i}$ to zero preserves detailed balance with respect to the same distribution $\boldsymbol{\pi}^\eq$, implying that the constant contribution $\lambda^0_{e\leftarrow i}$ to the mutual linearity relation, Eq. (2) of the main text, is 0.
Yet, since the susceptibility is defined by varying both rates $\boldsymbol{r}_i$
of the input edge $i$, 
we have in general $\lambda^1_{e\leftarrow i} \neq 0$ in 
$\jj_e(\r_{i}) = \lambda^1_{e\leftarrow i} \, \jj_i(\r_{i})$.
In that sense, such susceptibility characterizes the non-equilibrium response of the network.

In this Appendix, we show that the susceptibility presents a property of reciprocity 
that keeps track of detailed balance:
The susceptibility of output edge $e$ with respect to input edge $i$ has the same sign as
the susceptibility of output edge $i$ with respect to input edge $e$,
when they are defined from a reference dynamics where detailed balance holds.
Namely, starting from reference rates satisfying Eq.~\eqref{eq:defDB}, we define the corresponding
susceptibilities
$   \lambda^1_{e\leftarrow i}  $
and $   \lambda^1_{i\leftarrow e}  $
from
\begin{equation}
  \jmath_e(\boldsymbol{r}_i)
  = 
  \lambda^1_{e\leftarrow i} \,
  \jmath_i(\boldsymbol{r}_i)
\qquad
\text{and}
\qquad
  \jmath_i(\boldsymbol{r}_e)
  = 
  \lambda^1_{i\leftarrow e}\, 
  \jmath_e(\boldsymbol{r}_e)
  \:,
\end{equation}
assuming that $e$ and $i$ are not bridges.
We now show that these susceptibilities satisfy a symmetry that implies the property of reciprocity mentioned above.

The rate matrix is decomposed as 
$\mathbf R = - \mathbf S \mathbf V^\top$ 
where the $|\mathscr X|\times|\mathscr E|$ matrices $\mathbf S$  and $\mathbf V$ are defined in 
Eqs.~\eqref{eq:defS}-\eqref{eq:defV} and are the stoichiometric matrix and a weighted version of it. 
Detailed balance implies that
\begin{equation}
    \label{eq:VofSwhenDBholds}
    \mathbf V = \mathbf{\Pi}^{-1} \,\mathbf S\, \mathbf K
\end{equation}
where 
$\mathbf \Pi$ 
is a $|\mathscr X|\times|\mathscr X|$ diagonal matrix of elements
$\mathbf \Pi_{\x \x} = \pi^\eq_\x$ 
and
$\mathbf K$ 
is a $|\mathscr E|\times|\mathscr E|$ diagonal matrix of elements
\begin{equation}
\mathbf K_{ee} = 
    \pi^\eq_{\s(+e)} r_{+e} 
=
    \pi^\eq_{\s(-e)} r_{-e} 
=
    \big[
    \pi^\eq_{\s(+e)} 
    \pi^\eq_{\s(-e)} 
    r_{+e} 
    r_{-e} 
    \big]^{\frac 12}
\:.
\end{equation}
One checks indeed that the form of $\mathbf V$ given in Eq.~\eqref{eq:VofSwhenDBholds}
ensures that, for $\x\neq\x'$, 
$\mathbf R_{\x \x'}= r_{\pm e} $
if $\x=\s(\mp e)$ and $\x'=\s(\pm e)$ for some transition $e$
[and is zero otherwise]
for the matrix $\mathbf R = - \mathbf S \mathbf V^\top$.
(The diagonal elements are then automatically correct because $\mathbf 1$ is a left null vector of such a matrix $\mathbf R$).
Then, rewriting the expression of Eq.~\eqref{eq:lambda1limratiodet} of the susceptibility as
\begin{equation}
    \label{eq:lambda1limratiodetinverse}
   \lambda^1_{e\leftarrow i} 
   =
   \lim_{\si\to 0} 
    \frac
      {
       \langle e| 
       \big(\si \mathbf{1}+ \mathbf{V}^\top\, \mathbf{S} \big)^{-1}
       |i \rangle 
      }
      {
       \langle i| 
       \big(\si \mathbf{1}+ \mathbf{V}^\top\, \mathbf{S} \big)^{-1}
       |i \rangle 
      }
      \:,
\end{equation}
and using
$
\mathbf{V}^\top\, \mathbf{S} 
=
\mathbf K\, \mathbf S^\top \, \mathbf{\Pi}^{-1} \, \mathbf{S} 
=
\mathbf K\,
\big(\mathbf K\, \mathbf S^\top \, \mathbf{\Pi}^{-1} \, \mathbf{S} \big)^\top
\mathbf K^{-1}
=
\mathbf K\,
\big(\mathbf{V}^\top\, \mathbf{S}\big)^\top
\mathbf K^{-1}
$
together with Eq.~\eqref{eq:denom2.2},
we obtain
\begin{equation}
\frac{   \lambda^1_{e\leftarrow i} }{   \lambda^1_{i\leftarrow e} }
=
\frac
{\tau_{\del e}\mathbf{K}_{ee}}
{\tau_{\del i}\mathbf{K}_{ii}}
>0
\:.
\end{equation}
In summary, if the dynamics of a Markov chain satisfies detailed balance, the susceptibilities of an edge w.r.t.~the other and vice-versa have the same sign.

%------------------------------------------------
\sm{app:fig-details}{Details of the figures}
%------------------------------------------------

Figures can also be reproduced with the code publicly available at the repository~\cite{github}.

\medskip
\noindent
\textit{Figure 1}
\medskip

Below is represented the network used in Fig.~1 of the main text, with labeled nodes, and the rate matrix associated with it:

\

\begin{minipage}{.15\textwidth}
    \begin{center}
        \includegraphics[height=3.5cm]{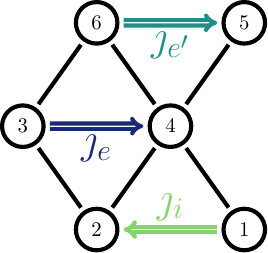}
    \end{center}
\end{minipage}%
\begin{minipage}{.8\textwidth}
    \begin{equation}
    \mathbf{R} =
    \left(
    \begin{array}{cccccc}
     -4.3-r_{+i} & r_{-i} & 0 & 3.9 & 0 & 0 \\
            r_{+i} & -5.7-r_{-i} & 3.7 & 2.1 & 0 & 0\\
            0 & 4.2 & -5.1 & 2.7 & 0 & 3.1\\
            4.3 & 1.5 & 0.2 & -12.5 & 3.1 & 2.5\\
            0 & 0 & 0 & 0.4 & -3.2 & 4.5\\
            0 & 0 & 1.2 & 3.4 & 0.1 & -10.1\\
    \end{array}
    \right)
\end{equation}
\end{minipage}

\

As represented above, the input edge is $i = 1 \to 2$, while the other edges considered are $e = 3 \to 4$ and $e' = 6 \to 5$. To obtain the currents, we find the stationary distribution $\boldsymbol{\pi}$ that uniquely satisfies $\mathbf{R} \boldsymbol{\pi} = 0$. In the top plot of Fig.~1 of the main text, we fix $r_{-i} =1$ and change $ r_{+i}$ in $[0,7]$. In the bottom plot, both input rates are chosen in the intervals $[0,3]$ at steps of 0.8, and the gray dashed lines are obtained by calculating affine coefficients when $r_{+i} = r_{-i} = 0$ and susceptibilities using Eq.~\eqref{eq:ratio}.

\medskip
\noindent
\textit{Figure 2}
\medskip

The multicyclic networks in Figure 2 were obtained from a Voronoi representation of two sets of random points in 2D. Respectively, we picked $50$ and $500$ random points on the unit square $[0,1] \times [0,1]$ to generate the network on the left and on the right of Figure 2. We then used the Python package Networkx to convert the Voronoi diagram into a directed graph and obtain the corresponding stoichiometric matrix $\mathbf{S}$. The resulting stoichiometric matrices have dimensions $74 \times 93$ for the left network and $447 \times 645$ for the right network. In both cases, we picked the transition rates as random real number from a uniform distribution $[0, 0.5]$ with working precision $\text{WP}=20$. We then built a numerical matrix $\mathbf{V}$ by plugging the numerical values of the rates into the definition Eq.~(\ref{eq:defV}). The rate matrix is then obtained via the expression $\mathbf{R} = - \mathbf{S}\mathbf{V}$ and used to compute the susceptibilities following the expressions of Appendix \ref{app:graphrepresentation}. In particular, combining Eqs.~(\ref{eq:lambda1limratiodet}, \ref{eq:denom2}, \ref{eq:lambda1limratiodetBISprime}, \ref{eq:towardsWstarwoodbury},\ref{eq:detSVstarfirstorder}) we compute the susceptibilities from the expression:
\begin{equation}\label{eq:susceptibilityinmathematica}
    \lambda^{1}_{e \leftarrow 1} = \frac{r_{+e}\sum_{\substack{\x\neq \star \\ \x\neq \s(+e)} }
\big( 
\operatorname{adj}\,
(-\mathbf{R}^\star_{\s (+e)})
\big)_{\x\x} 
- r_{-e}  \sum_{\substack{\x\neq \star \\ \x\neq \s(-e)} }
\big( 
\operatorname{adj}\,
(-\mathbf{R}^\star_{\s (-e)}) 
\big)_{\x\x}}{\sum_{\substack{\x}} \big( \operatorname{adj}\, (-\mathbf{R}_{\del 1})\big)_{\x\x}}
\end{equation}
Figure 2 shows the absolute value of the susceptibilities so obtained plotted using the Networkx package representation in Python. We refer to the supplementy files for an exemplification on how to code Eq.~(\ref{eq:susceptibilityinmathematica}) in Mathematica (see file tutorial.nb).

\medskip
\noindent
\textit{Figure 3}
\medskip

The rate matrix for the Myosin-V model used in the Appendix of the main text is obtained from Table S3 of the Supporting Information of \cite{mallory_kinetic_2020}. It is
\begin{equation}
    \mathbf{R} = 
    \left(
    \begin{array}{cccccc}
     \square & 0.49 & 0 & 0 & 302.8 & 0 \\
     10^5 & \square & 6.4 \times 10^4 & 0 & 0 & 302.8 \\
     0. & 4.6 \times 10^3 & \square & 4.64 \text{[Pi]} & 0. & 0. \\
     0. & 0. & 3 \times 10^3 & \square & 3.35 & 1.5 \times 10^{-2} \\
     0.62 & 0. & 0. & 15.2 & \square & 0. \\
     0. & 1.27 \times 10^{-6} & 0. & 6.9 \times 10^{-2} & 0. & \square \\
    \end{array}
    \right)
\end{equation}
where the diagonal elements can be obtained by forcing each column to sum to zero and [Pi] is measured in units of mM. The figure represents values of $\mathcal{J}_\text{ATP}$ and $\jj_\text{m}$ for concentrations [Pi] from 0 to 100mM at steps of 10, and the susceptibilities are obtained using Eq.~\eqref{eq:ratio}.

\end{document}